\documentclass[12pt, a4paper]{article}
\pdfoutput=1

\usepackage[latin1]{inputenc}
\usepackage{graphicx}
\usepackage{url}
\usepackage{amsmath}
\usepackage{amssymb}
\usepackage{ifpdf}
\newcommand{\R}[1]{\texttt{#1}}
\newcommand{\titleman}[1]{\begin{center}{\bf \LARGE
#1}\end{center}}

\newcommand{\bm}[1]{\mbox{\boldmath{$#1$}}}

\long\def\symbolfootnote[#1]#2{\begingroup%
\def\thefootnote{\fnsymbol{footnote}}\footnote[#1]{#2}\endgroup}

\usepackage{float}

\makeatletter
\newcommand\floatc@simplerule[2]{{\@fs@cfont #1} #2\par}
\newcommand\fs@simplerule{\def\@fs@cfont{\bfseries}\let\@fs@capt\floatc@simplerule
  \def\@fs@pre{\hrule height1.2pt depth0pt \kern4pt}%
  \def\@fs@mid{\vspace*{0.5em} \hrule height.3pt depth0pt \vspace*{0.8em} \kern4pt}%
  \def\@fs@post{\kern4pt \hrule height1.2pt depth0pt \kern4pt \relax}%
  \let\@fs@iftopcapt\iftrue}
\makeatother

\floatstyle{boxed}
\newfloat{algorithm}{thp}{alg}	
\floatname{algorithm}{Box}

\begin{document}

%\clearpage
%\pagestyle{empty}

\titleman{The Evolution of Boosting Algorithms \\ \vspace{.5cm}
\Large From Machine Learning to Statistical Modelling\symbolfootnote[1]{This article
is not an exact copy of the original published
article in \emph{Methods of Information in Medicine}. The definitive publisher-authenticated
version is available online (together with the companion review and an invited discussion) at: \url{http://dx.doi.org/10.3414/ME13-01-0122}.   \\
If citing, please refer to the original article: \\
 Mayr A, Binder H, Gefeller O, Schmid M. The Evolution of Boosting Algorithms -- From Machine Learning to Statistical Modelling. Methods Inf Med 2014; 53(6): 419--427.
}
}

%\begin{center}
%(part 1 of the review)  \\
%\end{center}

\begin{center}
Andreas Mayr\symbolfootnote[2]{\textit{Address for correspondence:}
Andreas Mayr, Institut f\"ur Medizininformatik, Biometrie und
Epidemiologie, Friedrich-Alexander Universit\"at
Erlangen-N\"urnberg, Waldstr. 6, 91054 Erlangen, Germany.}$^1$,
Harald Binder$^2$, Olaf Gefeller$^1$, \\ Matthias Schmid$^{1,
3}$ \vspace{0.5cm}

\begin{footnotesize}
 \footnotemark[1] Institut f\"ur Medizininformatik, Biometrie und Epidemiologie,\\
Friedrich-Alexander-Universit\"at Erlangen-N\"urnberg, Germany\\
\footnotemark[2] Institut f\"ur Medizinische Biometrie,
Epidemiologie und Informatik, \\ Johannes Gutenberg-Universit\"at Mainz, Germany \\
\footnotemark[3] Institut f\"ur Medizinische Biometrie, Informatik und Epidemiologie, \\ Rheinische Friedrich-Wilhelms-Universit\"at Bonn, Germany \\
\end{footnotesize}

\end{center}
\normalsize \vspace{1cm}

\begin{abstract}

%Boosting algorithms for statistical modelling have gained substantial interest in biomedical applications during the last decade. Although the initial idea of boosting was developed in the field of machine learning, nowadays boosting algorithms are often used to estimate unknown quantities in statistical models as well. Boosting methods have particular advantages in the presence of high-dimensional data that are frequently encountered in modern biomedical applications. In this review, we introduce to this evolving area by highlighting the historical evolution and fundamental principles of the different boosting approaches.

\noindent
\textbf{Background:} The concept of boosting emerged from the field of machine learning. The basic idea is to boost the accuracy of a weak classifying tool by combining various instances into a more accurate prediction. This general concept was later adapted to the field of statistical modelling. Nowadays, boosting algorithms are often applied to estimate and select predictor effects in statistical regression models.  \\
\noindent
\textbf{Objectives:}  This review article attempts to highlight the evolution of boosting algorithms from machine learning to statistical modelling. \\
\noindent
\textbf{Methods:} We describe the AdaBoost algorithm for classification as well as the two most prominent \textit{statistical boosting} approaches, gradient boosting and likelihood-based boosting for statistical modelling. We highlight the methodological background and present the most common software implementations.  \\
\noindent
\textbf{Results:} Although gradient boosting and likelihood-based boosting are typically treated separately in the literature, they share the same methodological roots and follow the same fundamental concepts. Compared to the initial machine learning algorithms, which must be seen as black-box prediction schemes, they result in statistical models with a straight-forward interpretation.  \\
\noindent
\textbf{Conclusions:} Statistical boosting algorithms have gained substantial interest during the last decade and
offer a variety of options to address important research questions in modern biomedicine.

\end{abstract}

%\clearpage
\clearpage
\section{Introduction}

Boosting algorithms represent one of the most promising methodological approaches for data analysis developed in the last two decades. The original algorithm \cite{nr:freund.schapire:1996} emerged from the field of machine learning, where it gained much interest and was soon considered as a powerful instrument to predict binary outcomes. The basic idea is to iteratively apply simple classifiers and to combine their solutions to obtain a better prediction result. The concept of boosting was later adapted to the field of statistical modelling, where it can be used to select and estimate the effect of predictors on a univariate response variable in different types of regression settings \cite{friedmanetal2000, friedman_2001}.

Following a recent focus theme on boosting algorithms in {\em Methods of Information in Medicine} \cite{editorial2012MIM}, the first aim of this review is to highlight the evolution of boosting from a black-box machine learning algorithm to a flexible tool to estimate and select interpretable statistical models. We will refer to this type of boosting algorithms as \textit{statistical boosting} algorithms. The second aim is to bridge the methodological gap between two different statistical boosting approaches which are typically treated separately in the literature, but share the same historical roots: gradient boosting \cite{BuhlmannHothorn06} and likelihood-based boosting \cite{TutzBinder}. Both are increasingly applied in biomedical settings for different kind of regression and prediction analysis \cite{faschingbauer2012macro, lin2013fast, SaiZhaFan2011}.

The reasons for the success of statistical boosting algorithms are (i) their ability to incorporate automated variable selection and model choice in the fitting process, (ii) their flexibility regarding the type of predictor effects that can be included in the final model and (iii) their stability in the case of high-dimensional data with possibly far more candidate variables than observations -- a setting where most conventional estimation algorithms for regression settings collapse. The application of boosting algorithms thus offers an attractive option for biomedical researchers: many modern biomedical settings like genome-wide association studies and research using other 'omics' technologies are specifically challenging regarding all three points mentioned above  \cite{li2005boosting, BinBenBul2013, mayr2014CI}.

This review is structured as follows: In Section~2, we introduce the machine-learning concept of boosting which led to the famous AdaBoost algorithm \cite{nr:freund.schapire:1996} for classification. In Section~3 we present the statistical view on boosting which paved the way for the development of statistical boosting algorithms that are suitable for general regression settings. We describe the generic algorithms for gradient boosting and likelihood-based boosting and present the most common software packages. In the concluding Section~4, we summarize the main findings and highlight the differences between AdaBoost and statistical boosting.

In a companion article \cite{boosting_part2}, we additionally document the significant progress in the methodological research on statistical boosting algorithms over the last few years.

\section{Boosting in machine learning}

The concept of boosting emerged from the field of supervised learning, which is the automated learning of an algorithm based on labelled data with observed outcome in order to make valid predictions for unlabelled future or unobserved data.  Supervised learning is a subdiscipline of {\em machine learning}, which also comprises unsupervised learning based on unlabelled data and semi-supervised learning which is a combination of supervised and unsupervised learning \cite{bishop2006pattern}. A supervised learning machine typically yields a generalization function $\hat{h}(\cdot)$ that provides the solution to a classification problem. The main goal of classification is to categorize objects into a  pre-defined set of classes.  For the remainder of this section we will consider the most common classification problem, where the outcome variable $Y$ has two classes, coded as $\{-1,1\}$. Note that this coding differs from the standard $\{0,1\}$ which is typically used in statistics for dichotomous outcomes.

The machine should learn  from a training sample $(y_1, \bm{x}_1), ..., (y_n, \bm{x}_n)$ with known class labels how to predict the class of a new observation $\bm{x}_{\text{new}}$. The predictors $\bm{x}_1,...,\bm{x}_n$ are realizations of $X$, and $n$ is the sample size. The task for the machine is to develop a prediction rule $\hat{h}(\cdot)$ to correctly classify a new observation:

\begin{eqnarray*}
 (y_1, \bm{x}_1),...,(y_n, \bm{x}_n)  \xrightarrow{\text{supervised learning}} \hat{h}(\bm{x}_{\text{new}})  = \hat{y}_{\text{new}}
\end{eqnarray*}

%Nowadays, modern machine-learning algorithms are often successfully applied in various fields of biomedical research \cite{malley2011statistical}. Applications include the areas of disease diagnostics \cite{sajda2006machine}, genome-wide association studies \cite{kruppa2012risk} or the estimation of patient-specific risks and probabilities \cite{malley2012probability, kruppaziegler2014}.

\subsection{The concept of boosting}

The success story of boosting began with a question, not with an algorithm. The theoretical discussion was if any weak learning tool for classification could be transformed to become also a strong learner \cite{KearnsValiant89}. In binary classification, a weak learner is defined to yield a correct classification rate at least slightly better than random guessing ($>$ 50\%). A strong learner, on the other hand, should be able to be trained to a nearly perfect classification (e.g., 99\% accuracy). This theoretical question is of high practical relevance as it is typically easy to construct a weak learner, but difficult to get a strong one \cite{ensemblesbook}. The answer, which laid the ground for the concept of boosting, is that any weak \emph{base-learner} can be potentially iteratively improved (\textit{boosted}) to become also a strong learner. To provide evidence for this concept, Schapire \cite{Schapire1989} and Freund \cite{Freund90} developed the first boosting algorithms.

Schapire and Freund later compared the general concept of boosting with \textit{``garnering wisdom from a council of fools''} \cite{boostingbook}. The ``fools'' in this case are the solutions of the simple base-learner: It classifies only slightly better than  the flip of a coin. A simple base-learner is by no means a practical classification rule, but even the simple base-learner must contain some valid information about the underlying structure of the problem. The task of a boosting algorithm is hence to learn from the iterative application of a weak learner and to use this information to combine it to an accurate classification.

However, just calling the weak learner multiple times on the same training sample would not change anything in its performance. The concept of boosting is not really to manipulate the base-learner itself to improve its performance but to manipulate the underlying training data by iteratively re-weighting the observations  \cite{boostingbook}. As a result, the base-learner in every iteration $m$ will find a new solution $\hat{h}^{[m]}(\cdot)$ from the data.

Via repeated application of the weak base-learner on observations that are weighted based on the base-learner's success in the previous rounds, the algorithm is forced to concentrate on objects that are hard to classify -- as observations that were misclassified before get higher weights. \textit{Boosting} the accuracy is achieved by \textit{increasing} the importance of ``difficult'' observations. In each iteration $m = 1,...,m_{\text{stop}}$, the weight vector $\bm{w}^{[m]} = (w_1^{[m]} ,...,w_n^{[m]})$ contains the individual weights of all observations depending on the success of their classification in previous iterations. During the iteration cycle, the focus is shifted towards observations that were misclassified up to the current iteration $m$.

In a final step, all previous results of the base-learner are combined into a more accurate prediction: The weights of better performing solutions of the base-learner are increased via an iteration-specific coefficient, which depends on the corresponding misclassification rate. The resulting weighted majority vote \cite{littlestone1989weighted} chooses the class most often selected by the base-learner while taking the error rate in each iteration into account (see point (5) in Box~\ref{alg:adaboost}).

This combination of forcing the algorithm to develop new strategies for problematic observations and rewarding the base-learner in the final aggregation for accurate solutions is the main idea of boosting. Following this concept, it can be shown that all weak learners can potentially be boosted to become also strong learners \cite{Schapire1989, Freund90}.

\subsection{AdaBoost}

The early boosting algorithms by Schapire \cite{Schapire1989} and Freund \cite{Freund90} were rather theoretical constructs for proving the idea of boosting than being suitable algorithms for practical usage.  However, they paved the way for the first concrete and -- still today -- most important boosting algorithm \textit{AdaBoost} \cite{nr:freund.schapire:1996}. AdaBoost was the first \emph{adaptive} boosting algorithm as it automatically adjusts its parameters to the data based on the actual performance in the current iteration: both the weights $w_i$ for re-weighting the data as well as the weights $\alpha_m$ for the final aggregation are re-computed iteratively. For a schematic overview, see Box~\ref{alg:adaboost} -- for worked out examples, we refer to \cite{ensemblesbook,boostingbook}.

\begin{algorithm}[t]
\begin{enumerate}
\item[] {\bf Initialization}

    \begin{enumerate}
    \item[(1)]  Set the iteration counter $m=0$ and the individual weights $w_i$ for observations $i = 1,...,n$ to $w_i^{[0]} = \frac{1}{n}$.
    \end{enumerate}

\item[]{\textbf{Base-learner}}

     \begin{enumerate}
       \item[(2)] Set $m := m + 1$ and compute the base-learner for the weighted data set:
       $$\text{re-weight observations with } w_1^{[m-1]},..., w_n^{[m-1]}   \xrightarrow{\text{base-learner}}   \hat{h}^{[m]}(\cdot) $$
      \end{enumerate}

\item[]{\textbf{Update weights}}

    \begin{enumerate}
     \item[(3)] Compute error rate and update the iteration-specific coefficient $\alpha_m$ $\rightarrow$ high values for small error rates. Update individual weights $w_i^{[m]}$ $\rightarrow$ higher values if observation was misclassified.
    \end{enumerate}

\item[]{\textbf{Iterate}}

    \begin{enumerate}
    \item[(4)] Iterate steps 2 and 3 until $m = m_{\text{stop}}$.
    \end{enumerate}

\item[]{\textbf{Final aggregation}}

    \begin{enumerate}
    \item[(5)] Compute the final classifier for a new observation $\bm{x}_{\text{new}}$:

    $$\hat{f}_{\text{Adaboost}}(\bm{x}_{\text{new}}) = \text{sign} \left(\sum_{m=1}^{m_{\text{stop}}} \alpha_{m} \hat{h}^{[m]}(\bm{x}_{\text{new}}) \right)$$
    \end{enumerate}

\end{enumerate}
\caption{ Schematic overview of the AdaBoost algorithm.}\label{alg:adaboost}
\end{algorithm}

%Many investigators concentrated on analyzing the performance of AdaBoost in empirical studies.

The introduction of AdaBoost gained much attention in the machine learning community.  In practice, it is often used with simple classification trees or stumps as base-learners and typically results in a dramatically improved performance compared to the classification by one tree or any other single base-learner \cite{Ridgeway99, meir2003introduction}. For example, Bauer and Kohavi \cite{Bauer99} report an average 27\% relative improvement in the misclassification error for AdaBoost compared with a single decision tree. The authors additionally compared the accuracy of AdaBoost with the one of Bagging \cite{Breiman96} in various settings. Bagging, in contrast to boosting, uses bootstrap generated samples to modify the training data and hence does not rely on the misclassification rate of earlier iterations. After their large-scale comparison, Bauer and Kohavi concluded that boosting algorithms, in contrast to Bagging, are able to reduce not only the variation in the base-learner's prediction error resulting from the use of different training data sets (variance), but also the average difference between predicted and true classes (bias). This view is also essentially supported by an analysis of Breiman \cite{Breiman98}. The success of AdaBoost allegedly led Breiman, who was a pioneer and leading expert in machine learning, to the statement \cite{elements2}: \textit{Boosting is the best off-the-shelf classifier in the world}.

\subsection{Overfitting}

A long-lasting discussion in the context of AdaBoost is its overfitting behavior. \textit{Overfitting} describes the common phenomenon that when a prediction rule concentrates too much on peculiarities of the specific sample of training observations it was optimized on, it will often perform poorly on a new data set \cite{dietterich1995overfitting}. To avoid overfitting, the task for the algorithm therefore should not be to find the best possible classifier for the underlying training sample, but rather to find the best prediction rule for a set of new observations.

The main control instrument to avoid overfitting in boosting algorithms is the stopping iteration $m_{\text{stop}}$. Very late stopping of AdaBoost may favor overfitting, as the complexity of the final solution increases. On the other hand, stopping the algorithm too early does not only inevitably lead to higher error on the training data but could as well result in a poorer prediction on new data (\textit{underfitting}). In the context of AdaBoost, it is nowadays consensus that although the algorithm may overfit \cite{Grove98, ratsch2001soft}, it often is quite resistent to overfitting \cite{BuhlmannHothorn06, ensemblesbook, boostingbook}.

In their initial article, Freund and Schapire \cite{nr:freund.schapire:1996} showed that the generalization error on a test data set of AdaBoost's final solution is bounded by the training error plus a term which increases with the number of boosting iterations and the complexity of the base-learner. This finding was apparently supported by the widely acknowledged principle known as Occam's Razor \cite{Blumer}, which roughly states that for predictions, more complex classifiers should be outperformed by less complex ones if both carry the same amount of information. However, this theoretical result is not supported by the observation that AdaBoost, in practice,  is often resistent to overfitting. As the complexity of the final AdaBoost solution depends mainly on the stopping iteration $m_{\text{stop}}$, following Occam's Razor, later stopping of the algorithm should yield poorer predictions \cite{ensemblesbook}.

One way to explain AdaBoost's overfitting behavior is based on the margin interpretation \cite{meir2003introduction,ratsch2001soft, Schapire98}: The margin of the final boosting solution, in brief, can be interpreted as the confidence in the prediction. With higher values of $m_{\text{stop}}$, this margin may still increase and lead to better predictions on the test data even if the training error is already zero \cite{Reyzin}. This theory was early questioned by results of Breiman \cite{Breiman99}, who developed the \emph{arc-gv} algorithm which should yield a higher margin than AdaBoost, but clearly failed to outperform it in practice with respect to prediction accuracy. Later, Reyzin and Schapire \cite{Reyzin} explained these findings with other factors like the complexity of the  base-learner. For more on the margin interpretation see the corresponding chapters in \cite{ensemblesbook,boostingbook}.

Another explanation of the -- seemingly contradictory -- results on the overfitting behavior of boosting is the use of the wrong performance criteria for evaluation (e.g., \cite{mease2008evidence}). The performance of AdaBoost has often been measured by evaluating the {\em correct classification rate}, and the resistance to overfitting has usually been demonstrated by focusing on this specific criterion only. However, the criterion that is optimized by AdaBoost is in fact not the correct classification rate but the so-called {\em exponential loss function}, and it can be shown that the two criteria are not necessarily optimized by the same predictions. For this reason some authors have argued that the overfitting behavior of AdaBoost should be analyzed by solely focusing on the exponential loss function \cite{BuhlmannHothorn06rej}. For example, B\"uhlmann and Yu \cite{BMW_comment_BYu} have provided empirical evidence that too large $m_{\text{stop}}$ can lead to overfitting regarding the exponential loss without affecting the misclassification rate.

\section{Statistical boosting}

Up to this point, we focused on the classical supervised learning problem where the task of boosting is to predict dichotomous outcomes. Nowadays, boosting algorithms are more often used to estimate the unknown quantities in general statistical models (\emph{statistical boosting}). In the remainder of this section, we will therefore broaden the scope and consider general regression settings where the outcome variable $Y$ can also be continuous or represent count data. The most important interpretation of boosting in this context is \emph{the statistical view} of boosting by Friedman et al.~\cite{friedmanetal2000}. It provided the basis for understanding the boosting concept in general and the success of AdaBoost in particular from a statistical point of view \cite{Ridgeway99} by showing that AdaBoost in fact fits an additive model.

Most solutions of machine-learning algorithms, including AdaBoost, must be seen as \textit{black-box} prediction schemes. They might yield very accurate predictions for future or unobserved data, but the way those results are produced and which role single predictors play are hardly interpretable. A statistical model, in contrast, aims at quantifying the relation between one or more observed predictor variables $\bm{x}$ and the expectation of the response $\text{E}(Y)$ via an \textit{interpretable} function $\text{E}(Y|X=\bm{x}) =  f(\bm{x})$. In cases of more than one predictor, the different effects of the single predictors are typically added, forming an additive model

\begin{eqnarray*}
 f(\bm{x}) = \beta_0 +  h_1(x_1) + \cdots + h_p(x_p)
\end{eqnarray*}

where $\beta_0$ is an intercept and $h_1(\cdot)$,...,$h_p(\cdot)$ incorporate the effects of predictors $x_1, ..., x_p$, which are components of $X$. The corresponding model class is called generalized additive models ('GAM', \cite{hastietib}) and the aim is to model the expected value of the response variable, given the observed predictors via a link-function $g(\cdot)$:

\begin{eqnarray*}
 g(\text{E}(Y|X=\bm{x})) = \beta_0 + \sum_{j=1}^p h_j(x_j)
\end{eqnarray*}

GAMs are by definition no black boxes but contain \textit{interpretable} additive predictors: The partial effect of predictor $x_1$, for example, is represented by $h_1(\cdot)$. The direction, the size and the shape of the effect can be visualized and interpreted  -- this is a main difference towards many tree-based machine learning approaches.

The core message delivered with the statistical view of boosting is that the original AdaBoost algorithm with regression-type base-learners (e.g., linear models, smoothing splines), in fact, fits a GAM for dichotomous outcomes via the exponential loss in a stage-wise manner. The work by Friedman et al.~\cite{friedmanetal2000} therefore provided the link between a successful machine-learning approach and the world of statistical modelling \cite{Ridgeway99}.

\subsection{Gradient boosting}

The concept of the statistical view of boosting  was further elaborated by Friedman \cite{friedman_2001} who presented a boosting algorithm optimizing the empirical risk via steepest gradient descent in function space. Generally, the optimization problem for estimating the regression function $f(\cdot)$ of a statistical model, relating the predictor variables $X$ with the outcome $Y$, can be expressed as

\begin{eqnarray*}
 \hat{f} (\cdot)   =  \underset{f(\cdot)}{\operatorname{argmin}} \Bigg\{  \text{E}_{Y,X} \Big[ \rho(Y, f(X)) \Big] \Bigg\} ,
\end{eqnarray*}

where $\rho(\cdot)$ denotes a loss function. The most common loss function is the $L_2$ loss $\rho(y,f(\cdot)) =  (y - f(\cdot))^2$, leading to classical least squares regression of the mean: $f(\bm{x}) = \text{E} (Y | X = \bm{x})$. In practice, with a learning sample of observations $(y_1, \bm{x}_1), ...,(y_n, \bm{x}_n)$ we minimize the \textit{empirical} risk:

\begin{eqnarray*}
 \hat{f} (\cdot)   =  \underset{f(\cdot)}{\operatorname{argmin}} \left\{ \frac{1}{n} \sum_{i=1}^n  \rho(y_i, f(\bm{x}_i)) \right\}
\end{eqnarray*}

The fundamental idea of gradient boosting is to fit the base-learner not to re-weighted observations, as in AdaBoost, but to the negative gradient vector $\bm{u}^{[m]}$ of the loss function $\rho(y, \hat{f}(x))$ evaluated at the previous iteration $m-1$:

$$ \bm{u}^{[m]} = \left( u^{[m]}_i \right)_{i=1,...,n} = \left( -  \left. \frac{\partial }{\partial f} \rho(y_i,f )\right|_{ f = \hat{f}^{[m-1]}(\cdot)} \right)_{i=1,...,n}  $$

In case of the $L_2$ loss, $\rho(y,f(\cdot)) =  \frac{1}{2} (y - f(\cdot))^2$ leads simply to re-fitting the residuals $y - f(\cdot)$. In every boosting iteration $m$, the base-learner is hence directly fitting the errors made in the previous iteration $y - f(\cdot)^{[m-1]}$. Keeping this principle in mind, it becomes obvious that both AdaBoost and gradient boosting follow the same fundamental idea: Both algorithms \textit{boost} the performance of a simple base-learner by iteratively shifting the focus towards problematic observations that are `difficult' to predict. With AdaBoost, this shift is done by up-weighting observations that were misclassified before. Gradient boosting identifies difficult observations by large residuals computed in the previous iterations.

\begin{algorithm}[ht!]
\begin{enumerate}

\item[] {\bf Initialization}
    \begin{enumerate}

    \item[(1)]  Set the iteration counter $m=0$. Initialize the additive predictor $\hat{f}^{[0]}$ with a starting value, e.g. $\hat{f}^{[0]} := (\bm{0})_{i=1,...,n}$.  Specify a set of base-learners $h_1(x_1),..., h_p(x_p)$.

    \end{enumerate}

\item[]{\textbf{ Fit the negative gradient}}

 \begin{enumerate}
   \item[(2)] Set $m := m + 1$.
   \item[(3)] Compute the negative gradient vector $\bm{u}$ of the loss function evaluated at the previous iteration:

  $$ \bm{u}^{[m]} = \left( u^{[m]}_i \right)_{i=1,...,n} = \left( -  \left. \frac{\partial }{\partial f} \rho(y_i,f )\right|_{ f = \hat{f}^{[m-1]}(\cdot)} \right)_{i=1,...,n}  $$

 \item[(4)] Fit the negative gradient vector $\bm{u}^{[m]}$ separately to every base-learner:

 $$ \bm{u}^{[m]} \xrightarrow{\rm base-learner}
\hat{h}_j^{[m]}(x_j) \quad \text{for } j=1,...,p.$$
\end{enumerate}

\item[]{\textbf{Update one component}}

\begin{enumerate}
\item[(5)] Select the component $j^*$ that best fits the negative gradient vector:

    $$ j^* = \underset{1 \leq j \leq p}{\operatorname{argmin}}\sum_{i=1}^n (u_{i}^{[m]} - \hat{h}_{j}^{[m]}(x_j))^2 \ . $$

\item[(6)] Update the additive predictor $\hat{f}$ with this component

$$ \hat{f}^{[m]}(\cdot) = \hat{f}^{[m-1]} (\cdot) + \text{sl}
\cdot \hat{h}_{j^*}^{[m]}(x_{j^*}) \ , $$

where sl is a small step length $(0 < \text{sl} \ll 1)$. A typical value in practice is 0.1.

\end{enumerate}

\item[]{\textbf{Iteration}}

\begin{enumerate}
\item[] Iterate steps (2) to (6) until $m=m_{\text{stop}}$. \\
\end{enumerate}

\end{enumerate}
\caption{Component-wise gradient boosting algorithm}\label{alg:gradboost}
\end{algorithm}

Generally, the underlying base-learner can be any regression technique; the most simple base-learner is a classical linear least-squares model with $h(\bm{x}) = \bm{x}^{\top}\beta$. If $\bm{x}$ is assumed to have a non-linear effect on the response, smoothing splines could be used \cite{friedman_2001}. B\"uhlmann and Yu \cite{BuehlmannYu2003} further developed the gradient boosting approach by applying \emph{component-wise} smoothing splines as base-learners. The fundamental idea is that different predictors are fitted by separate base-learners $h_j(\cdot)$, $j=1,...,p$.  Typically, each base-learner $h_j(\cdot)$ corresponds to one component $x_j$ of $X$ and in every boosting iteration (as proposed in \cite{friedman_2001}) only a small amount of the fit of the best-performing base-learner is added to the current additive predictor. The authors demonstrated that the resulting algorithm in combination with the $L_2$ loss outperforms classical additive modelling in terms of prediction accuracy. This approach was further developed by B\"uhlmann \cite{Buehlmann2006} who specially focused on high-dimensional data settings.

B\"uhlmann and Hothorn \cite{BuhlmannHothorn06} gave an overview of gradient boosting algorithms from a statistical perspective presenting a generic functional gradient descent algorithm (see Box~\ref{alg:gradboost}).  As in \cite{friedman_2001}, base-learners are used to fit the negative gradient vector of the corresponding loss function. The algorithm descends the empirical risk via steepest gradient descent in function space, where the function space is provided by the base-learners.  Each base-learner typically includes one predictor and in every boosting iteration only the best-performing base-learner and hence the best performing component of $X$ is included in the final model. This procedure effectively leads to data-driven variable selection during the model estimation. The base-learners $h_1(x_1),..., h_p(x_p)$ reflect the type of effect the corresponding components will contribute to the final additive model, which offers the same interpretability as any other additive modelling approach. Examples for base-learners can be trees as in classical boosting algorithms, but commonly simple regression tools like linear models or splines are used to include linear as well as non-linear effects on the response. Generally, it is consensus in the literature that base-learners should be \emph{weak} in the sense that they do not offer too complex solutions in a single iteration (e.g., \textit{penalized} splines with small degrees of freedom \cite{Schmid:Hothorn:boosting-p-Splines}). %However, there is no formal definition for \textit{weak} or \textit{strong} learners in the context of statistical boosting.

In contrast to standard estimation methods, component-wise gradient boosting also works for high dimensional data where the number of predictors exceeds the number of observations ($p>n$). Furthermore, it is relatively robust in cases of multicollinearity. Due to the small step length in the update step (a typical value is 0.1 \cite{mboostTut}) in combination with early stopping (Section 3.3), gradient boosting incorporates shrinkage of effect estimates in the estimation process: The absolute size of the estimated coefficients is intentionally reduced -- this is a similarity to penalized regression approaches as the Lasso \cite{Tibsh96}. Shrinkage of effect estimates leads to a reduced variance of estimates and should therefore increase the stability and accuracy of predictions \cite{elements2}.

The gradient boosting approach can be used to optimize any loss function that is at least convex and differentiable: The framework is specifically not restricted to statistical distributions that are members of the exponential family as in classical GAMs. For example, Ma and Huang \cite{ma2005} applied gradient boosting with an adapted ROC (receiver operating characteristics) approach, optimizing the area under the ROC curve for biomarker selection from high-dimensional microarray data.

\subsection{Likelihood-based boosting}

When considering statistical models, estimation in low-dimensional settings typically is performed by maximizing a likelihood. While such a likelihood can also be used to define a loss function in gradient boosting, a boosting approach could also be built on base-learners that directly maximize an overall likelihood in each boosting step. This is the underlying idea of likelihood-based boosting, introduced by Tutz and Binder \cite{TutzBinder}. When the effects of the predictors $x_1,\ldots,x_p$ can be specified by a joint parameter vector $\beta$, the task is to maximize the overall log-likelihood $l(\beta)$. Given a starting value or estimate from a previous boosting step $\hat\beta$, likelihood-based boosting approaches use base-learners for estimating parameters $\gamma$ in a log-likelihood $l(\gamma)$ that contains the effect of $\hat\beta$ as a fixed offset. For obtaining small updates, similar to gradient boosting, a penalty term is attached to $l(\gamma)$. The estimates $\hat\gamma$ are subsequently used to update the overall estimate $\hat\beta$. For continuous response regression models, including an offset is the same as fitting a model to the residuals from the previous boosting step, and maximization of $l(\gamma)$ by a base-learner becomes standard least-squares estimation with respect to these residuals. In this special case, likelihood-based boosting thus coincides with gradient boosting for $L_2$ loss \cite{BuehlmannYu2003}.

\begin{algorithm}[t]
\begin{enumerate}

\item[] {\bf Initialization}
    \begin{enumerate}

    \item[(1)]  Set the iteration counter $m=0$. Initialize the additive predictor $\hat{f}^{[0]}$ with a starting value, e.g. $\hat{f}^{[0]} := (\bm{0})_{i=1,...,n}$ or the maximum likelihood estimate $\hat\beta_0$ from an intercept model (if the overall regression model includes an intercept term).
    \end{enumerate}

\item[]{\textbf{Candidate models}}

 \begin{enumerate}
   \item[(2)] Set $m := m + 1$.
   \item[(3)] For each predictor $x_j$, $j=1, ...,p$ estimate the corresponding functional term $\hat{h}_j(\cdot)$, as determined by parameter $\gamma_j$, by attaching a penalty term to the log-likelihood $l(\gamma_j)$, which includes $\hat{f}^{[m-1]}(\cdot)$ as an offset.

\end{enumerate}

\item[]{\textbf{Update one component}}

\begin{enumerate}
\item[(4)] Select the component $j^*$ that results in the candidate model with the largest log-likelihood $l(\hat\gamma_{j^*})$:

    $$ j^* = \underset{1 \leq j \leq p}{\operatorname{argmax}} \  l(\hat\gamma_{j}) $$

\item[(5)] Update $\hat{f}^{[m]}$ to

$$ \hat{f}^{[m]}(\cdot) = \hat{f}^{[m-1]}(\cdot) +  \hat{h}_{j^*}^{[m]}(x_{j^*}) \ , $$
potentially adding an intercept term from maximum likelihood estimation.

\end{enumerate}

\item[]{\textbf{Iteration}}

\begin{enumerate}
\item[] Iterate steps (2) to (5) until $m=m_{\text{stop}}$. \\
\end{enumerate}

\end{enumerate}
\caption{Component-wise likelihood-based boosting algorithm}\label{alg:Likeboost}
\end{algorithm}

Component-wise likelihood-based boosting performs variable selection in each step, i.e. there is a separate base-learner for fitting a candidate model for each predictor $x_j$ by maximizing a log-likelihood $l(\gamma_j)$. The overall parameter estimate $\hat\beta$ then only is updated for that predictor $x_{j^*}$ which results in the candidate model with the largest log-likelihood $l(\hat\gamma_j)$. In linear models, $\gamma_j$ is a scalar value, and the penalized log-likelihood takes the form $l(\gamma_j) - \lambda\gamma_j^2$, where $\lambda$ is a penalty parameter that determines the size of the updates. Component-wise likelihood-based boosting then generalizes stagewise regression \cite{efronetal}.

For a schematic overview of component-wise likelihood-based boosting see Box~\ref{alg:Likeboost}. Tutz and Binder \cite{TutzBinder} applied this principle to generalized additive models with B-spline base-learners. Likelihood-based boosting for generalized linear models was introduced in another article by Tutz and Binder \cite{TutBin2007} and an approach for generalized additive mixed models was described by Groll and Tutz \cite{GroTut2012}. In these approaches, the best component for an update is selected according to the deviance in each boosting step. To decrease the computational demand with a large number of covariates, the likelihood-based boosting approach for the Cox proportional hazards model \cite{BinSch2008} instead uses a score statistic.

While component-wise likelihood-based boosting often provides results similar to gradient boosting (e.g., \cite{BinSch2008}), the use of standard regression models in the boosting steps allows for adaptation of techniques developed for the standard regression setting. For example, unpenalized covariates can be incorporated in a straightforward way by not incorporating these into the penalty term attached to $l(\gamma)$ \cite{BinSch2008}, but estimating their parameters together with a potential intercept term in steps (1) and (5). Approximate confidence intervals for the estimated covariate effects can be obtained by combining hat matrices from the individual boosting steps \cite{TutzBinder}.

\subsection{Early stopping of statistical boosting algorithms}

Although there are different influential factors for the performance of boosting algorithms, the stopping iteration $m_{\text{stop}}$ is considered to be the main tuning parameter \cite{Mayr_mstop}. Stopping the algorithm before its convergence (\textit{early stopping}) prevents overfitting (Section 2.3) and typically improves prediction accuracy. In case of statistical boosting, $m_{\text{stop}}$ controls both shrinkage of effect estimates and variable selection. The selection of $m_{\text{stop}}$ hence reflects the common bias-variance trade-off in statistical modelling: Large values of $m_{\text{stop}}$ lead to more complex models with higher variance and small bias. Smaller values of $m_{\text{stop}}$ lead to sparser models with less selected variables, more shrinkage and reduced variance \cite{Mayr_mstop}.

To prevent overfitting, it is crucial not to consider the stopping iteration $m_{\text{stop}}$ that leads to the best model on the training data but to evaluate the effect of $m_{\text{stop}}$ on separate test data. If no additional data are available, two general approaches are commonly applied:

The first is to use information criteria (AIC, BIC or gMDL \cite{hansen2001model}) which evaluate the likelihood on training data but additionally penalize too complex models by adding a multiple of their degrees of freedom. There are two problems with this approach: (i) for component-wise boosting algorithms these information criteria rely on an estimation of the degrees of freedom that is known to underestimate the true values \cite{hastiecomment}; (ii) they are only available for a limited number of loss functions.

The second, more general approach is to apply resampling or cross-validation techniques to subsequently divide the data into test and training sets and choose $m_{\text{stop}}$ by evaluating the models on the test data. For the evaluation, it is crucial to use the same loss function the algorithm aims to optimize. If the algorithm in a binary classification setting optimizes the exponential loss, one should use the exponential loss and not the misclassification rate to select $m_{\text{stop}}$. The optimal $m_{\text{stop}}$ is hence the one which leads to the smallest average empirical loss on the out-of-sample test data.

\subsection{Implementation and computational complexity}

Most implementations of statistical boosting algorithms are included in freely available add-on packages for the open source programming environment R \cite{r-core:2012}. Worked out examples and R-code for applying the most important implementations are provided in the Appendix of this article.

Gradient boosting is implemented in the add-on package \textbf{mboost} (\textit{model-based boosting}, \cite{pkg:mboost:CRAN:2.1}). The package provides a large variety of pre-implemented loss functions and base-learners yielding wide-ranging possibilities for almost any statistical setting where regression models can be applied. For an overview of how \textbf{mboost} can be used in biomedical practice, see Hofner et al.~\cite{mboostTut}. An alternative implementation of gradient boosting is provided with the \textbf{gbm} package \cite{gbm} which focuses on trees as base-learners. Likelihood-based boosting for generalized linear and additive regression models is provided by the add-on package \textbf{GAMBoost} \cite{GAMBoost} and an implementation of the Cox model is contained in the package \textbf{CoxBoost} \cite{CoxBoost}.

One  of the main advantages of statistical boosting approaches compared to standard estimation schemes is that they are computationally feasible in $p>n$ situations. The computational complexity of statistical boosting approaches depends mainly on the number of separate base-learners. In case of component-wise boosting, the complexity increases linearly with $p$ \cite{BuehlmannYu2006}. The computationally most burdensome part of applying statistical boosting in practice is the selection of the stopping iteration $m_{\text{stop}}$. In case of applying information criteria (as the AIC), this involves multiplication of $n \times n$ matrixes for each boosting iteration, which becomes computationally problematic for data settings with large $n$. The computing-time to select $m_{\text{stop}}$ via resampling procedures depends mainly on $m_{\text{stop}}$ itself, the number of resamples $B$ and $p$ \cite{Mayr_mstop}. In practice, selecting $m_{\text{stop}}$ via resampling can be drastically fastened by applying parallel computing, which is implemented in all R packages for statistical boosting.

\section{Conclusion}

%In this article we highlighted the concept and evolution of boosting, which is arguably one of the most important methodological contributions to the field of machine learning in the last decades.

%The introduction of AdaBoost was a milestone for the development of purely data-driven prediction rules. It was, however, the \textit{statistical view of boosting} \cite{friedmanetal2000} which paved the way for the success story of boosting algorithms in statistical modelling and their application in biomedical research. In contrast to AdaBoost, statistical boosting does not necessarily focus on classification problems but can be applied to various type of regression settings.

One reason for the success of statistical boosting algorithms is their straight-forward interpretation. While competing machine learning approaches (including AdaBoost) may also yield accurate predictions in case of complex data settings, they must be seen as \textit{black boxes}: The structure of the underlying data is considered irrelevant and the way different predictors contribute to the final solution remains unknown.  Statistical boosting algorithms, in contrast, are typically applied with simple regression-type functions as base-learners and therefore yield classical statistical models, reflecting the contribution of different predictors on an outcome variable of interest. As a result, their solution offers the same interpretation as any other model in classical regression analysis -- only that it was derived by applying one of the most powerful prediction frameworks available in the toolbox of a modern statistician.

We presented two specific frameworks for statistical boosting: gradient boosting and likelihood-based boosting. Although both algorithms are typically treated separately in the literature, both follow the same structure and share the same historical roots. In some special cases like the $L_2$ loss and Gaussian response they coincide. While gradient boosting is a more general approach and also allows for distribution-free regression settings like optimizing a ROC curve \cite{ma2005} or boosting quantile regression \cite{Fens2011}, likelihood-based boosting carries the advantage that it delivers the Hessian matrix, which can be used to compute approximate confidence intervals for the estimated predictor effects.

It is by no means an exaggeration to forecast that the application of statistical boosting algorithms in biomedical research will increase in the years to come. One of the main reasons for this development is that the number of candidate variables and predictors for modern biomedical research has continuously been increasing in recent years. In this type of settings, statistical boosting algorithms can demonstrate their full strengths via automated variable selection and model choice while still providing the same interpretability most biomedical research relies on.

\subsection*{Acknowledgements}

The work on this article was supported by the Deutsche Forschungsgemeinschaft (DFG) (\texttt{www.dfg.de}), grant SCHM 2966/1-1.

\subsection*{Discussion}

An invited discussion on this article and its companion review \cite{boosting_part2} can be found in the same issue of \textit{Methods of Information in Medicine} as the original article: \\

B\"uhlmann P, Gertheiss J, Hieke S,  Kneib T, Ma S, Schumacher M, Tutz G, Wang CY, Wang Z, Ziegler A. Discussion of ``The Evolution of Boosting Algorithms'' and ``Extending Statistical Boosting''. Methods Inf Med 2014; 53: XX-XX.

\footnotesize
\bibliography{bibliography_boosting_AM}

\normalsize
\section*{Appendix}

In the main article, we highlighted the concept and evolution of boosting, which is arguably one of the most important methodological contributions to the field of machine learning in the last decades. The introduction of AdaBoost was a milestone for the development of purely data-driven prediction rules. It was, however, the \textit{statistical view of boosting} \cite{friedmanetal2000} which paved the way for the success story of boosting algorithms in statistical modelling and their application in biomedical research. In contrast to AdaBoost, statistical boosting does not necessarily focus on classification problems but can be applied to various type of regression settings.

This Appendix provides examples on how to apply statistical boosting algorithms in practice. For the gradient boosting approach, we will concentrate on the implementation provided by the R \cite{r-core:2012} add-on package \textbf{mboost} (\textit{model-based boosting}, \cite{pkg:mboost:CRAN:2.1}).  For the likelihood-based boosting approach, we present examples and Code for the implementations provided by \textbf{GAMBoost} \cite{GAMBoost} as well as the \textbf{CoxBoost} \cite{CoxBoost} package. All packages and the underlying open source programming environment R are freely available at \url{http://r-project.org}.

\subsection*{Gradient boosting}

The \textbf{mboost} package provides a large number of pre-implemented loss functions (families) and base-learners which can be combined by the user  yielding wide-ranging possibilities for almost any statistical  setting where regression models can be applied. For a detailed tutorial describing the application of \textbf{mboost}, including also tables presenting the various base-learners and families, see \cite{mboostTut}.

\subsubsection*{The \texttt{bodyfat} data}

The overall aim of this application is to compute accurate predictions for the
body fat of women based on available anthropometric measurements. Observations
of 71 German women are available with the data set provided by Garcia et al. \cite{garcia}
and included in \textbf{mboost}. This illustrative example has been already used for demonstration purposes in the context of boosting \cite{BuhlmannHothorn06, mboostTut}.

The response variable is the body fat measured by DXA (\texttt{DEXfat}) which can be
seen as the gold standard to measure body fat. However, DXA measurements are too
expensive and complicated for a broad use. Anthropometric measurements as waist
or hip circumferences are in comparison very easy to measure in a standard
screening. A prediction formula only based on these measures could therefore be
a valuable alternative with high clinical relevance for daily usage. In total, the data set contains 8 continuous variables as possible predictors.

In the original publication \cite{garcia}, the presented prediction formula was
based on a linear model with backward-elimination for variable selection. The
resulting final model utilized hip circumference (\texttt{hipcirc}), knee breadth
(\texttt{kneebreadth}) and a compound covariate (\texttt{anthro3a}) which is defined as
the sum of the logarithmic measurements of chin skinfold, triceps skinfold and
subscapular skinfold:

\begin{center}
\begin{small}
\begin{verbatim}
R> library(mboost) ## load package
R> data(bodyfat)   ## load data
R>
R> ## Reproduce formula of Garcia et al., 2005
R> lm1 <- lm(DEXfat ~ hipcirc  + kneebreadth + anthro3a, data = bodyfat)
R> coef(lm1)
 (Intercept)     hipcirc kneebreadth    anthro3a
 -75.2347840   0.5115264   1.9019904   8.9096375
\end{verbatim}
\end{small}
\end{center}

A very similar model can be easily fitted by statistical boosting, applying
\texttt{glmboost()} (which uses linear base-learners) with default settings:

\begin{small}
\begin{verbatim}
R> ## Estimate same model by glmboost
R> glm1 <- glmboost(DEXfat ~ hipcirc  + kneebreadth + anthro3a,
+                   data = bodyfat)
R> coef(glm1, off2int = TRUE)    ## off2int adds the offset to the intercept
 (Intercept)     hipcirc kneebreadth    anthro3a
 -75.2073365   0.5114861   1.9005386   8.9071301
\end{verbatim}
\end{small}

Note that in this case we used the default family \texttt{Gaussian()} leading to boosting with the $L_2$ loss.

\subsubsection*{Different loss functions}

The loss function can be easily adapted via the \texttt{family} argument inside the fitting functions of mboost.  For example, we now fit the median (with \R{family = Laplace()}),  apply Gamma regression (\R{GammaReg()}) and Huber's loss for robust regression (\R{Huber()}). To compare the prediction accuracy, we leave out the first ten observations as test data: \\

\begin{small}
\begin{verbatim}
R> ## separate training and test data
R> dat.train <- bodyfat[-(1:10),-1] ## removing age
R> dat.test  <- bodyfat[1:10,-1]
R>
R> ## original formula
R> lm1 <- glm(DEXfat ~ hipcirc + kneebreadth + anthro3a,
R>            data =dat.train)
R> ## boosting:
R> ## "response ~ ." includes all remaining variables in the candidate model
R> glm1 <- glmboost(DEXfat ~ . , data = dat.train) ## L_2 -> Gaussian
R> glm2 <- glmboost(DEXfat ~ . , data = dat.train, family = Laplace())
R> glm3 <- glmboost(DEXfat ~ . , data = dat.train, family = GammaReg())
R> glm4 <- glmboost(DEXfat ~ . , data = dat.train, family = Huber())
R>
R> ## predictions on test data: mean squared error of prediction
R> mean((predict(lm1, dat.test) - dat.test$DEXfat)^2)       ## orig.
[1] 8.782721
R> mean((predict(glm1, dat.test) - dat.test$DEXfat)^2)      ## boosting
[1] 5.141709
R> mean((predict(glm2, dat.test) - dat.test$DEXfat)^2)      ## median
[1] 19.02454
R> mean((exp(predict(glm3, dat.test)) - dat.test$DEXfat)^2) ## gamma
[1] 8.748015
R> mean((predict(glm4, dat.test) - dat.test$DEXfat)^2)      ## robust
[1] 5.234016
\end{verbatim}
\end{small}

Not surprisingly, the best performing loss function (regarding the mean squared error of prediction on the 10 test observations) is the $L_2$ loss leading to classical regression of the mean. Note, that in this small illustrative example, we used the default of 100 boosting iterations with linear base-learners.

\subsubsection*{Different base-learners}

Up to now, we used the fitting function \R{glmboost()} which automatically applies linear ordinary least squares base-learner. Via the function \R{gamboost()} one can use basically the same formula interface, however alternatively P-spline base-learners are fitted for all variables.

\begin{small}
\begin{verbatim}
R> # now with P-splines
R> gam1 <- gamboost(DEXfat ~ hipcirc + kneebreadth + anthro3a,
+                   data = dat.train)
R> mean((predict(gam1, dat.test) - dat.test$DEXfat)^2)
[1] 9.274717
R> ## show partial effects
R> par(mfrow=c(1,3)) ## three plots in a row
R> plot(gam1)
\end{verbatim}
\end{small}

The \R{plot()} function for an object of the class \R{gamboost} automatically displays the partial effects of the different covariates on the response.

\begin{center}
\includegraphics[width = \textwidth]{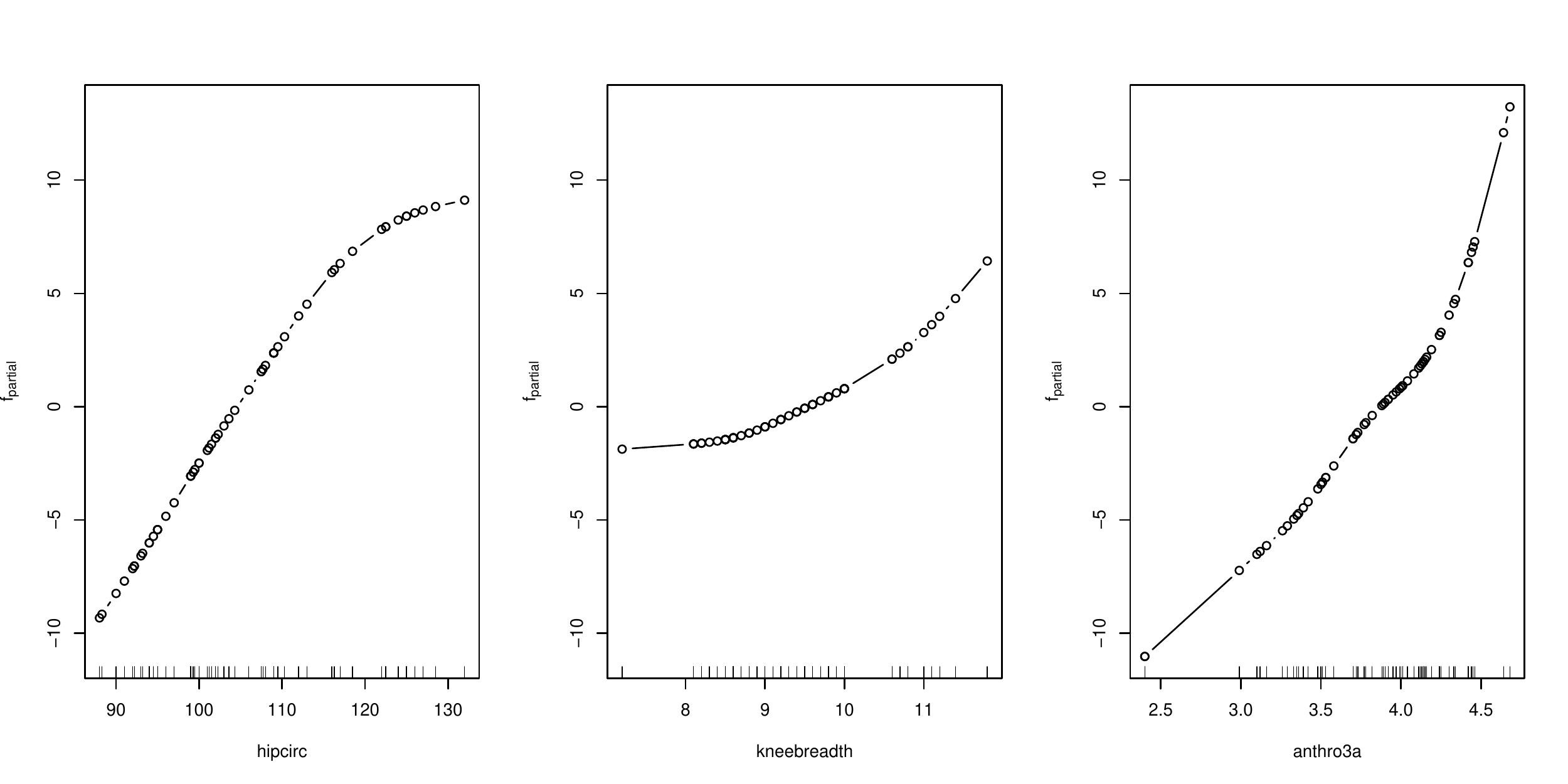}\\
\end{center}

\begin{small}
\begin{verbatim}
R> # with all variables as candidates
R> gam2 <- gamboost(DEXfat ~  . ,  data = dat.train)
R> mean((predict(gam2, dat.test) - dat.test$DEXfat)^2)
[1] 4.931875
R>
R> # Try the same with different family -> median
R> gam3 <- gamboost(DEXfat ~  . , family = Laplace(),  data = dat.train)
R> mean((predict(gam3, dat.test) - dat.test$DEXfat)^2)
[1] 10.58769
\end{verbatim}
\end{small}

However, it is of course also possible to include some variables with a linear effect and others with smooth effects. This can be done by specifying the type of base-learner inside the formula interface. For example, \R{bols(x1)} applies an ordinary least squares base-learner for variable \R{x1}, while \R{bbs(x2)} incorporates a P-splines base-learner (cubic P-splines with second order differences, 20 inner knots and 4 degrees of freedom) for variable \R{x2}. For more on boosting with splines, see \cite{Schmid:Hothorn:boosting-p-Splines}.

\begin{small}
\begin{verbatim}
R> # linear effect for hipcirc, smooth effects for kneebreadth and anthro3a
R>
R> gam4 <- gamboost(DEXfat ~ bols(hipcirc) + bbs(kneebreadth) + bbs(anthro3a),
+                   data = dat.train)
R>
R> mean((predict(gam4, dat.test) - dat.test$DEXfat)^2)
[1] 9.725139
R> par(mfrow=c(1,3)) ## three plots in a row
R> plot(gam4)
\end{verbatim}
\end{small}

\begin{center}
\includegraphics[width = \textwidth]{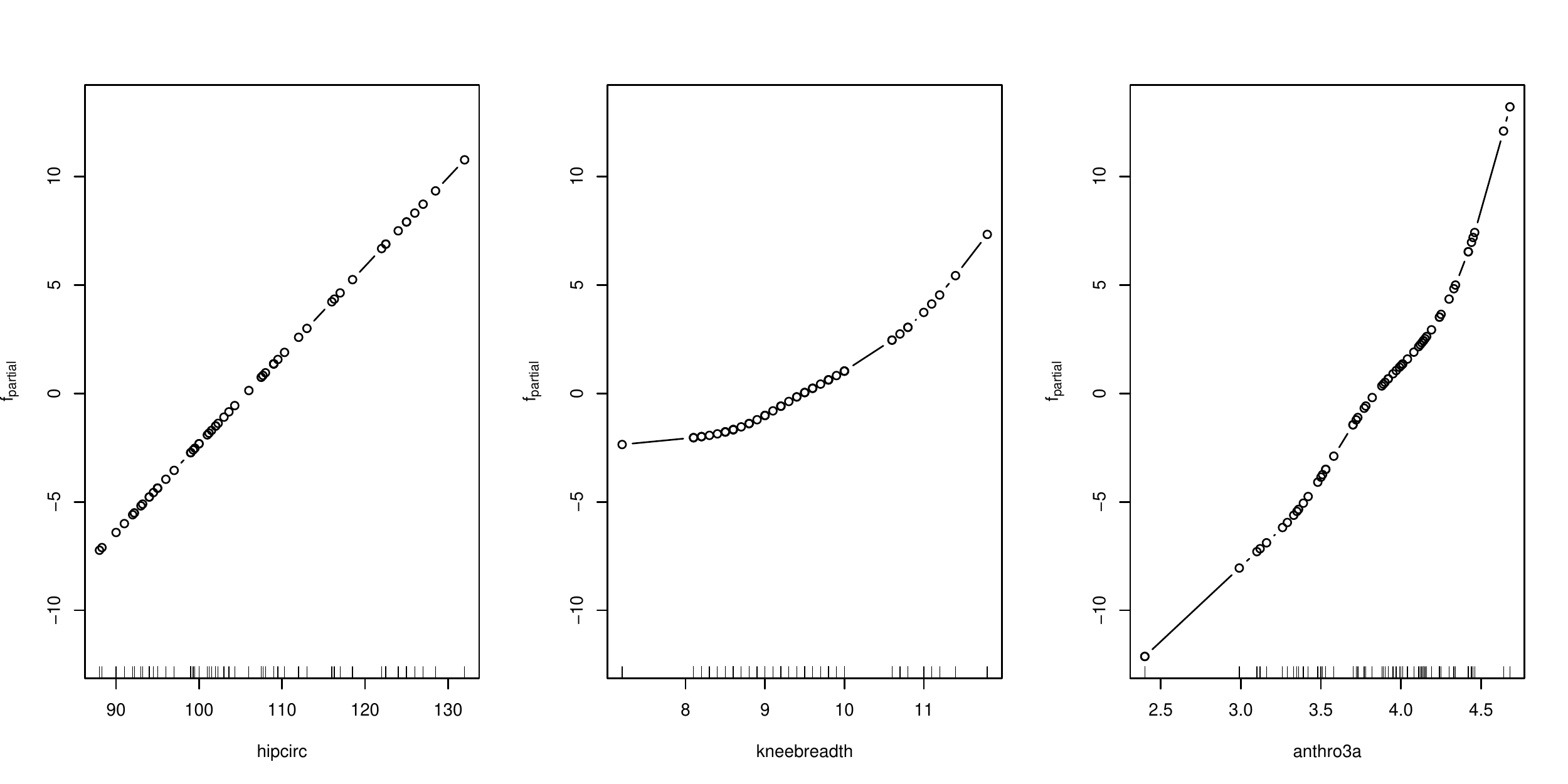}\\
\end{center}

Compared to the partial effects of \R{gam1}, the model now included a linear effect for \R{hipcirc}.

\subsubsection*{Early stopping}

The main tuning parameter of boosting algorithms is the stopping iteration. In \textbf{mboost}, the number of boosting iterations can be specified inside the \R{boost\_control()} function which can then be passed along to the fitting function \R{glmboost()} and \R{gamboost()}. The default number of boosting iterations is \R{mstop = 100}.

\begin{small}
\begin{verbatim}
R> ## now mstop = 500
R> gam1 <- gamboost(DEXfat ~ hipcirc + kneebreadth + anthro3a, data = dat.train,
+                   control = boost_control(mstop = 500, trace = TRUE))
[   1] ...................................................... -- risk: 523.347
[ 126] ...................................................... -- risk: 476.3329
[ 251] ...................................................... -- risk: 456.9567
[ 376] ......................................................
Final risk: 445.9373
\end{verbatim}
\end{small}

Once a model is fitted, the number of boosting iterations can also be modified by simple indexing:

\begin{small}
\begin{verbatim}
R> ## go back to mstop = 450
R> gam1 <- gam1[450]
R> mean((predict(gam1, dat.test) - dat.test$DEXfat)^2)
[1] 15.68723
\end{verbatim}
\end{small}

As described in the article, to select the optimal stopping iteration one can either apply information criteria as the AIC or resampling procedures. For both general approaches there exist pre-implemented functions in \textbf{mboost}:

\begin{small}
\begin{verbatim}
R> AIC(gam1)
[1] 3.552281
Optimal number of boosting iterations: 149
Degrees of freedom (for mstop = 149): 9.593731
R> gam1 <- gam1[149]
R> mean((predict(gam1, dat.test) - dat.test$DEXfat)^2)
[1] 10.51821
\end{verbatim}
\end{small}

Note that \textbf{mboost} per default applies a bias-corrected version of the AIC, which was proposed by Hurvich et al. \cite{Hurvich98}. However, the main problem with those criteria is that they (i) are not available for all loss functions, and more importantly, (ii) in case of boosting rely on estimations of the degrees of freedom which are severely biased. As a result, we generally think that resampling or cross-validation techniques in combination with the empirical loss are more appropriate to determine the stopping iteration \cite{Mayr_mstop}. In \textbf{mboost}, the function \R{cvrisk()} can be applied to automatically select the best-performing stopping iteration. It can be used to carry out cross-validation, subsampling or bootstrapping (the default). Note that in case of heavily unbalanced data sets, concerning a dichotomous outcome or an important predictor, it may be necessary to apply stratified resampling techniques via the argument \R{strata}. When the package \textbf{parallel} is available, \R{cvrisk()} applies parallel computing (if the machine contains more than one core).

\begin{small}
\begin{verbatim}
R> set.seed(123)
R> cvr <- cvrisk(gam1) ## default: 25-fold Bootstrap (takes a few seconds)
R> plot(cvr)
\end{verbatim}
\end{small}

\begin{center}
\includegraphics[scale=.5]{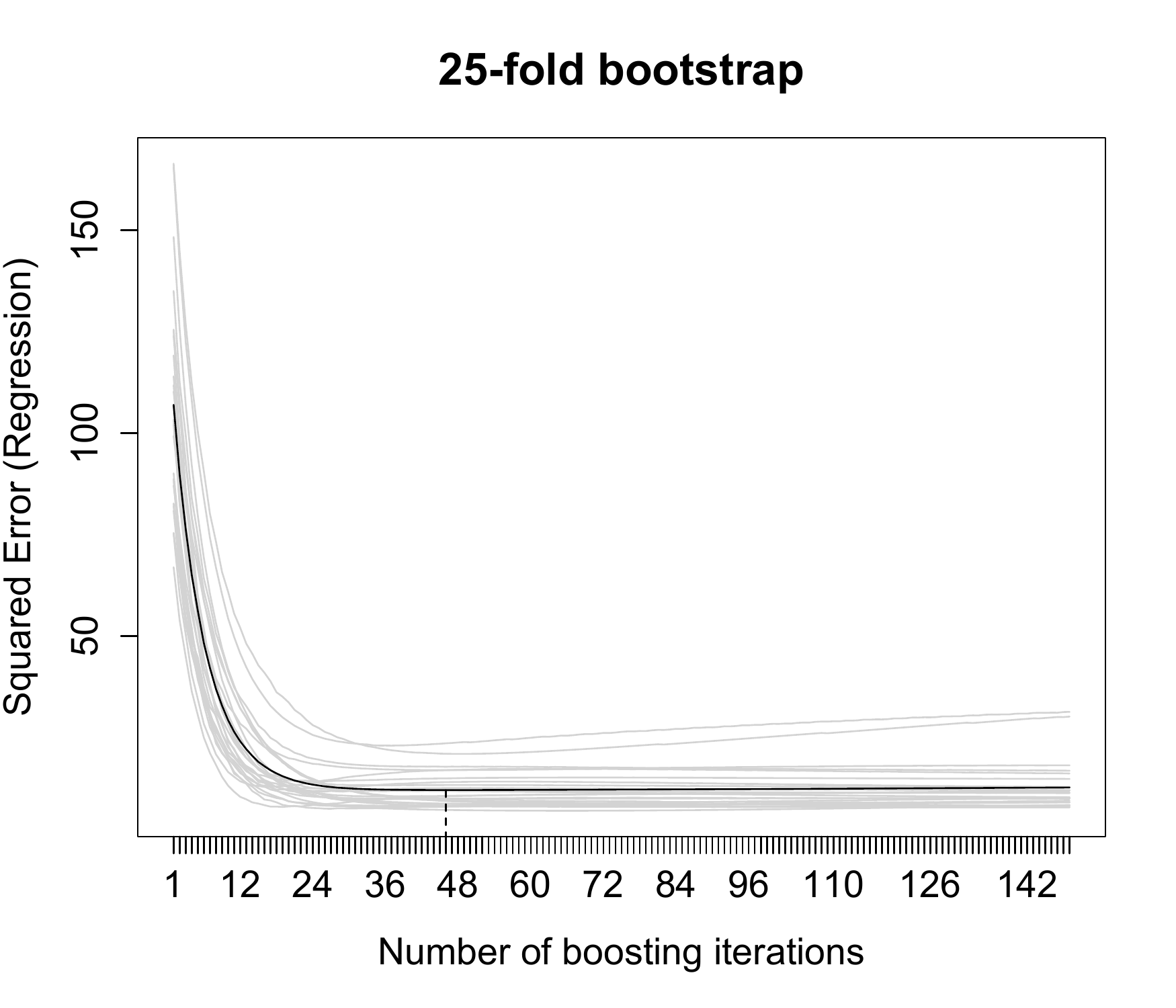}\\
\end{center}

\begin{small}
\begin{verbatim}
R> mstop(cvr) ## distract the optimal mstop
[1] 46
R> mean((predict(gam1[46], dat.test) - dat.test$DEXfat)^2)
[1] 7.924422
\end{verbatim}
\end{small}

Now we led the component-wise boosting algorithm select the most important predictors:

\begin{small}
\begin{verbatim}
R> ## include all variables, 100 iterations
R> gam2 <- gamboost(DEXfat ~ . , data = dat.train)
R> cvr <- cvrisk(gam2) ## 25-fold bootstrap
R> mstop(cvr)
[1] 38
R> gam2 <- gam2[mstop(cvr)] ## set to optimal iteration
R> mean((predict(gam2, dat.test) - dat.test$DEXfat)^2)
[1] 3.901965
\end{verbatim}
\end{small}

\subsubsection*{Early stopping controls smoothness of splines}

The stopping iteration of boosting algorithms does not only control the variable selection properties, but also the amount of shrinkage and the smoothness of effect estimates. The function \R{gamboost()} per default implements cubic P-splines with second order differences, 20 inner knots and 4 degrees of freedom. However, as the same spline base-learner can be chosen and updated in various iterations (and the final solution is the sum of those base-learner effects), boosting can adapt to an arbitrarily higher-order smoothness and complexity \cite{Schmid:Hothorn:boosting-p-Splines, BuehlmannYu2003}. This will be demonstrated in a small simulated example.

\begin{small}
\begin{verbatim}
R> set.seed(1234)
R> x <- runif(150, -0.2, 0.2)
R> y = (0.5 - 0.9* exp(-50*x^2))*x + 0.02 *rnorm(150)
R> y <- y[order(x)]  ## order obs by size of x
R> x <- x[order(x)]  ## just for easier plotting
\end{verbatim}
\end{small}

We simulated a clearly non-linear effect of the predictor \R{x} on the response \R{y}. We now fit step-by-step a simple univariate model via a P-spline base-learner and the classical $L_2$ loss while plotting the model fit at the different iterations. Furthermore, we control how boosting is re-fitting the residual reducing the empirical loss (sum of squares of residuals).

\begin{small}
\begin{verbatim}
R> par(mfrow = c(1,2)) ## two plots in one device
R>
R> ## model fit
R> plot(x, y, las = 1, main = "model fit at m = 1" ) ## observations
R> curve((0.5 - 0.9* exp(-50*x^2))*x, add=TRUE, from = -.2,
+         to = .2,  lty = 2,  lwd = 2)               ## true function
R> ## now carry out one boosting iteration
R> gam1 <- gamboost(y ~ x, control = boost_control(mstop = 1))
R> lines(x , fitted(gam1), col = 2, lwd = 2)         ## plot fitted values
R>
R> ##  residual plot
R> plot(x, y - fitted(gam1[1]) , ylab = "residuals", main = "residuals at m = 1",
+       ylim = c(-.1, .1), las = 1)                  ## residuals
R> lines(smooth.spline(x, y - fitted(gam1)),
+        col = 4, lwd = 2)                           ## show remaining structure
 \end{verbatim}
\end{small}

\begin{center}
 \includegraphics[width = \textwidth]{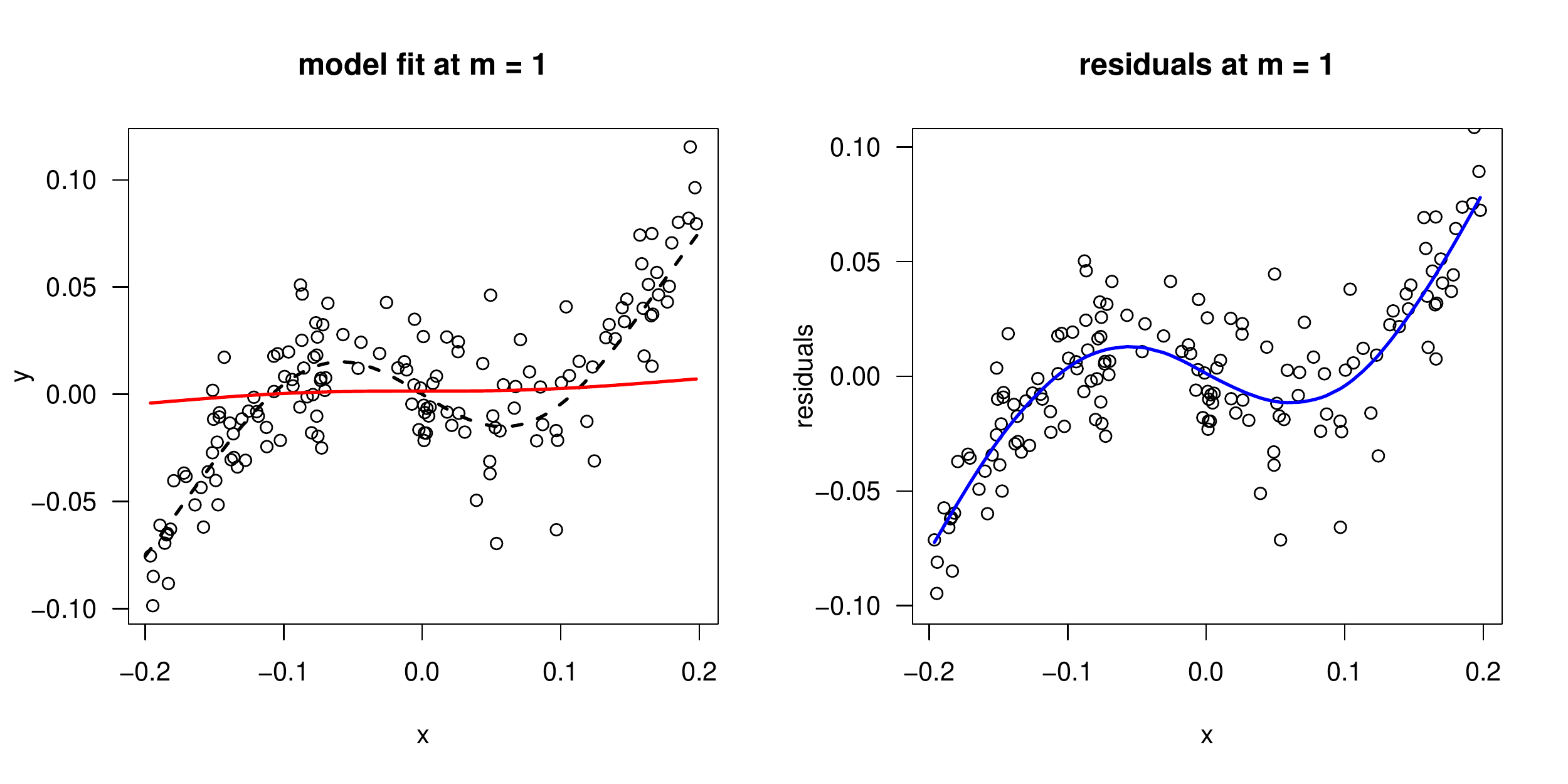}
\end{center}

Now we repeat the same after 5 iterations. Note that we make use of the possibility to modify the number of iterations of an existing model by simple indexing \R{gam1[5]}.

\begin{small}
\begin{verbatim}
R> ## model fit
R> plot(x, y, las = 1, main = "model fit at m = 5" )
R> curve((0.5 - 0.9* exp(-50*x^2))*x, add=TRUE, from = -.2, to = 0.2,
+         lty =2,  lwd = 2)
R> lines(x , fitted(gam1[5]), col = 2, lwd = 2)
R>
R> ## residual plot
R> plot(x, y - fitted(gam1[5]) , ylab = "residuals", main = "residuals at m = 5",
+       las = 1,  ylim = c(-.1, .1))
R> lines(smooth.spline(x, y - fitted(gam1[5])), col = 4, lwd = 2)
 \end{verbatim}
\end{small}

\begin{center}
 \includegraphics[width = \textwidth]{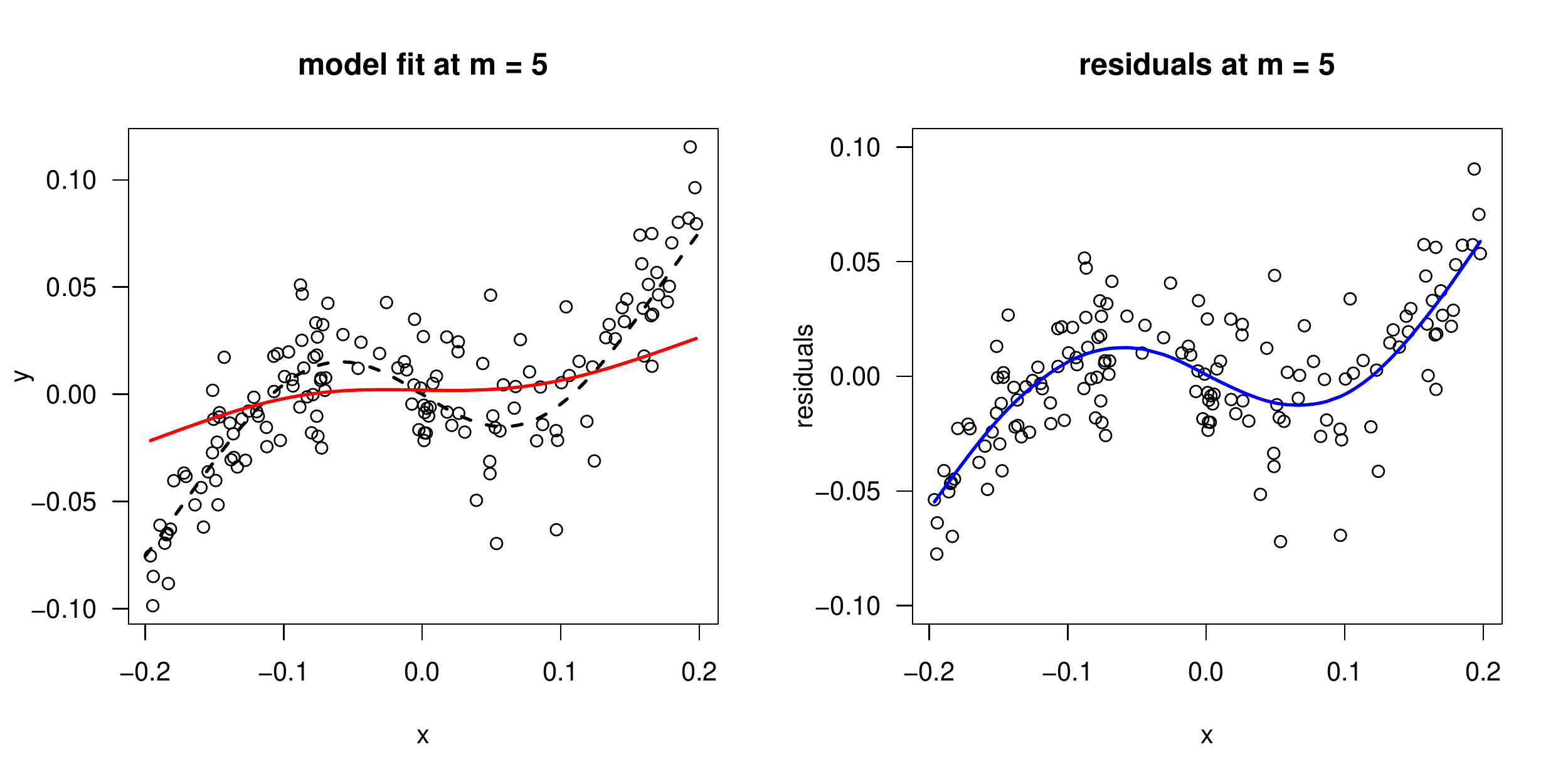}
\end{center}

We then repeat this procedure for iterations \R{mstop = 30} and \R{mstop = 50}. One can clearly observe how the model fit (red line) adapts nicely to the true function (dashed line) and how the structure in the residual (blue line) disappears.

\begin{center}
 \includegraphics[width = \textwidth]{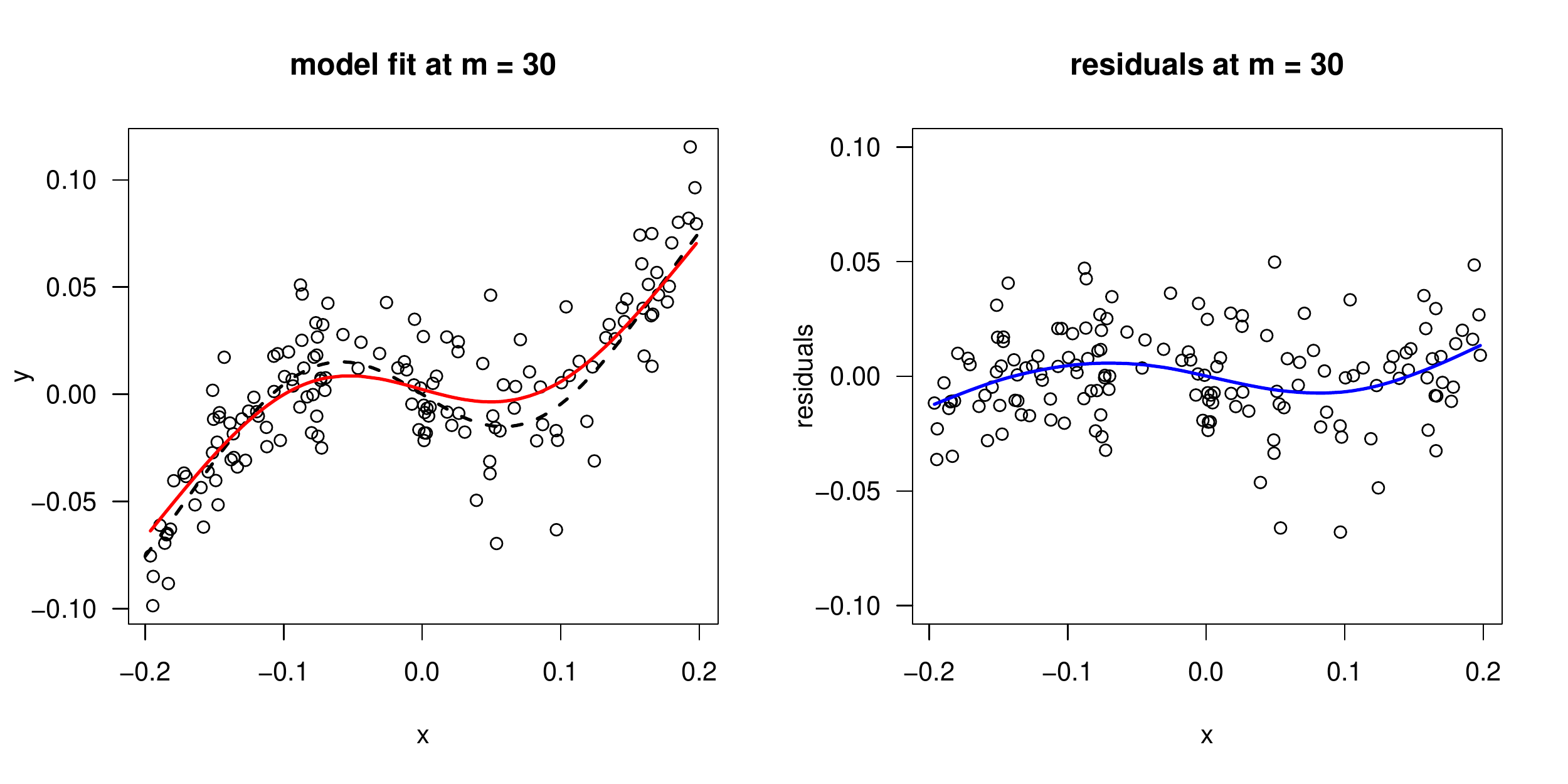}\\
  \includegraphics[width = \textwidth]{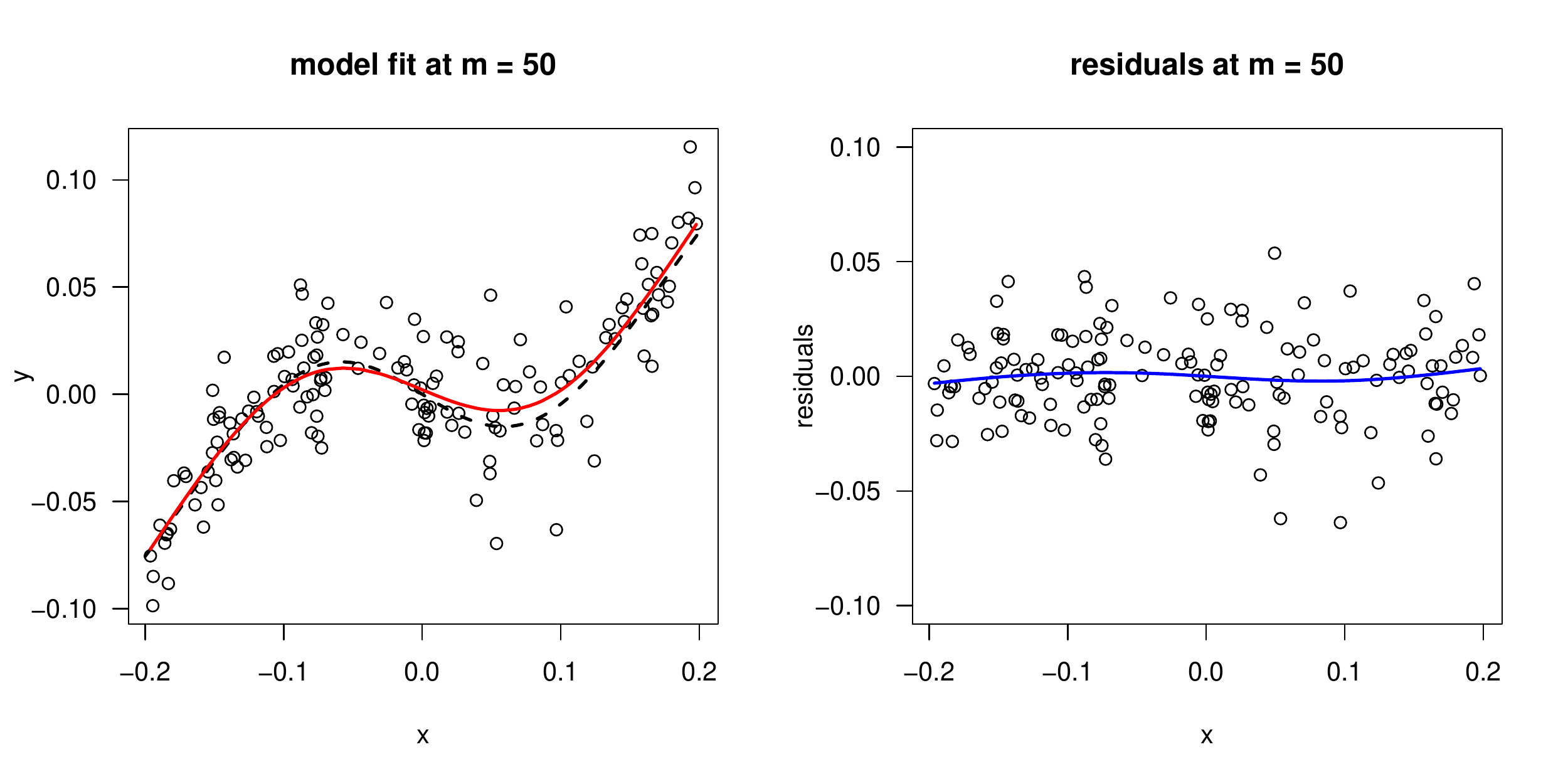}
\end{center}

To determine the optimal stopping iteration, we again perform 25-fold bootstrapping and choose the iteration that optimizes the predictive risk on the out-of-bootstrap observations (as implemented in \R{cvrisk()}).

\begin{small}
    \begin{verbatim}
R> set.seed(123)
R> cvr <- cvrisk(gam1, grid = 1:200) ## max mstop = 200
R> mstop(cvr)
[1] 110
    \end{verbatim}
\end{small}

\includegraphics[width = \textwidth]{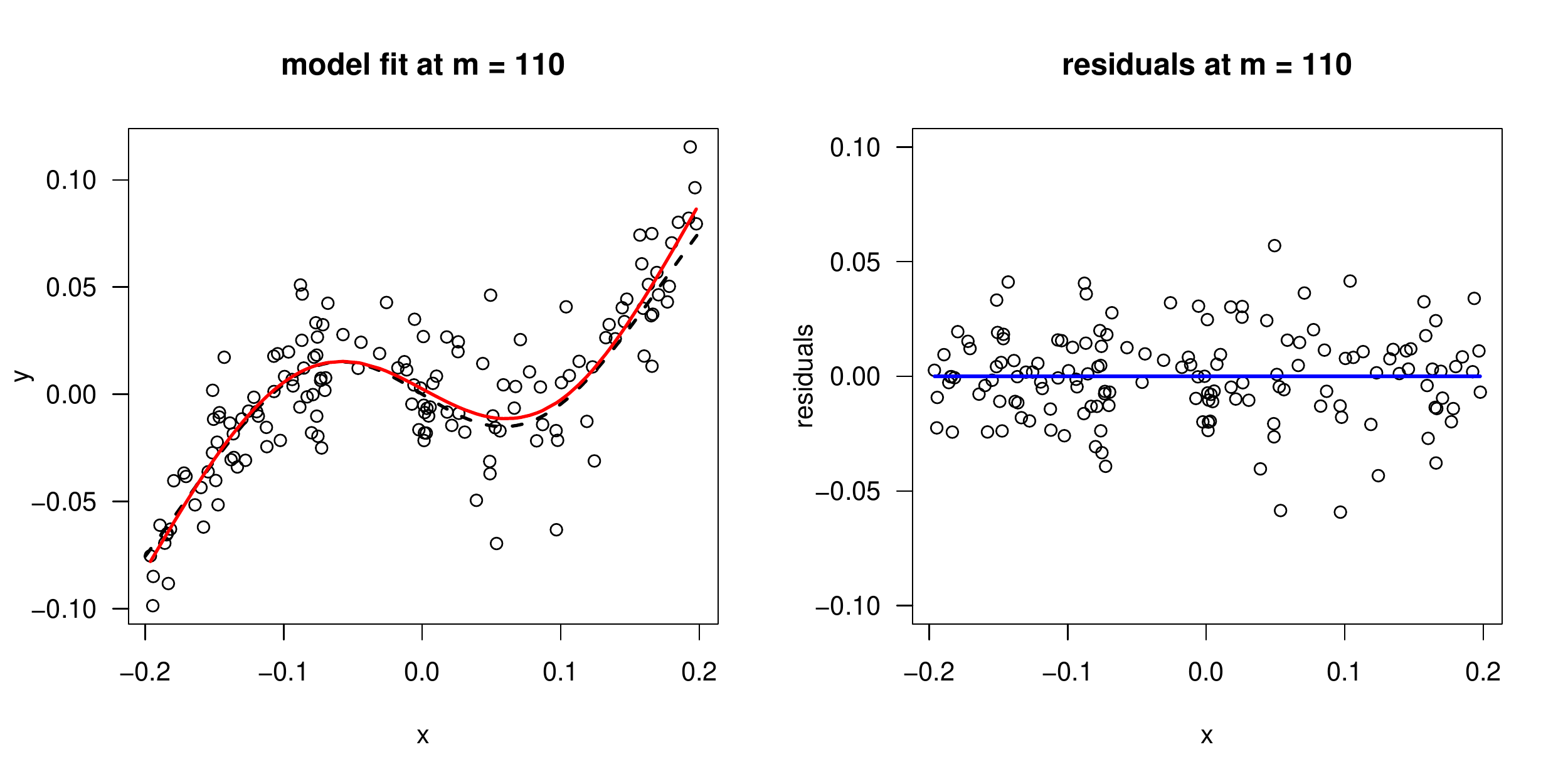}

\subsubsection*{Slow overfitting behavior}

Although the selection of the stopping iteration \R{mstop} is crucial to control variable selection, shrinkage and the overall smoothness of spline base-learners, it has been stated that statistical boosing algorithms show rather \textit{slow overfitting behavior} \cite{BuhlmannHothorn06}. This can also be shortly demonstrated based on the simple simulation from above. While the optimal number of boosting iterations was \R{mstop = 110} one could also analyse what happens if the algorithm is stopped much later. \\

\begin{small}
\begin{verbatim}
R> ## model fit mstop = 1000
R> plot(x, y, las = 1, main = "model fit at m = 1000" )
R> curve((0.5 - 0.9* exp(-50*x^2))*x, add=TRUE, from = -.2, to = 0.2,
+         lty =2,  lwd = 2)
R> lines(x , fitted(gam1[1000]), col = 2, lwd = 2)
R>
R> ## model fit mstop = 50000
R> plot(x, y, las = 1, main = "model fit at m = 50000" )
R> curve((0.5 - 0.9* exp(-50*x^2))*x, add=TRUE, from = -.2, to = 0.2,
+         lty =2,  lwd = 2)
R> lines(x , fitted(gam1[50000]), col = 2, lwd = 2)
 \end{verbatim}
\end{small}

We therefore compare the model fit for \R{mstop = 1000} iterations and the one resulting from  \R{mstop = 50000} with the optimal model: The plotted curve for \R{mstop = 1000} looks basically the same as the optimal one -- although we used about 10 times as much iterations as necessary. This is a nice indication of the rather slow overfitting properties of boosting. However, with \R{mstop = 50000} it gets clear that also boosting will eventually overfit -- the resulting curve is definitely much to rough.

\includegraphics[width = \textwidth]{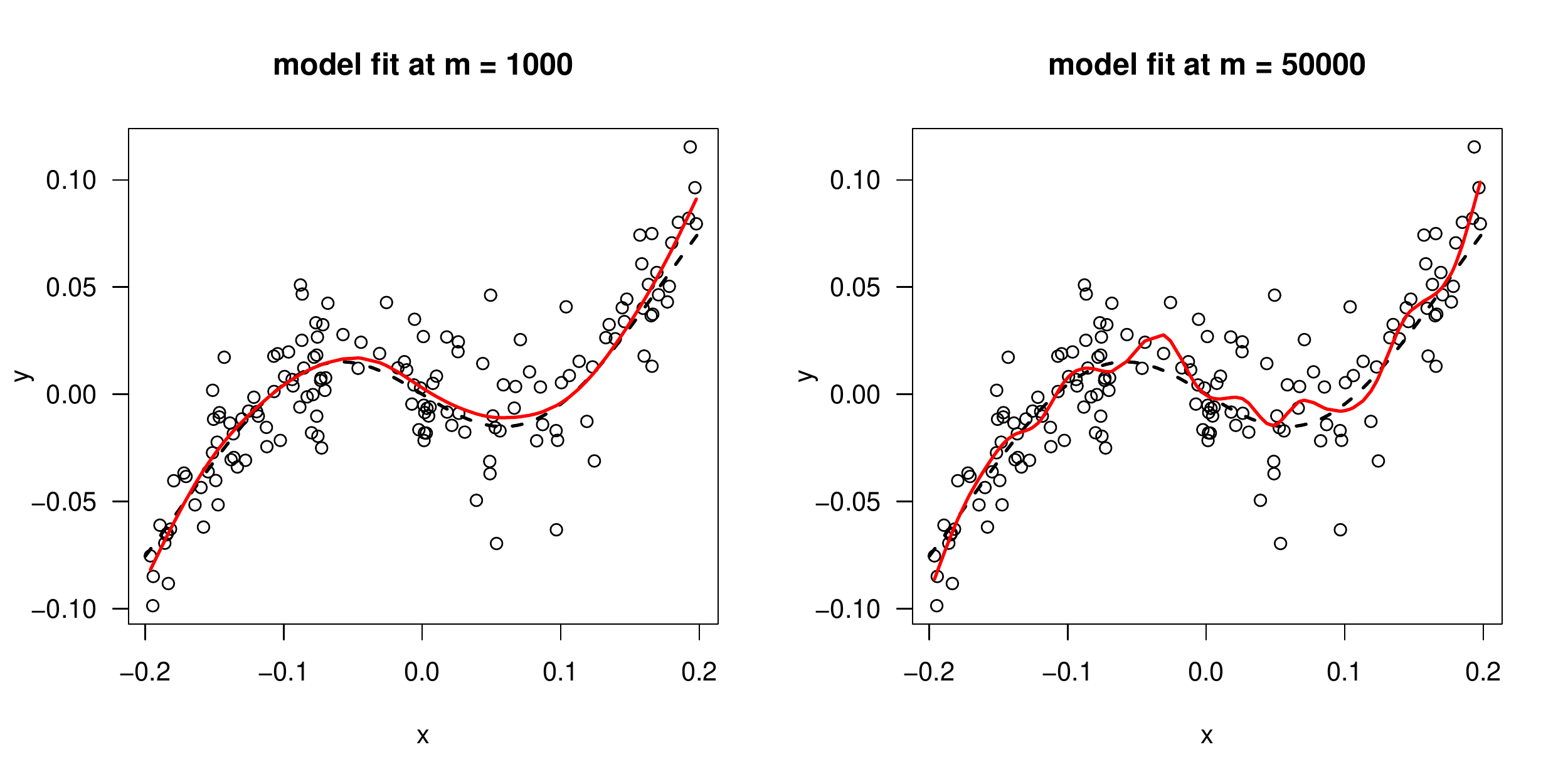}

\subsection*{Likelihood-based boosting}

The application of likelihood-based boosting, as implemented in the \textbf{GAMBoost} package is in fact not very different from gradient boosting in \textbf{mboost}.

\subsubsection*{The \texttt{bodyfat} data}

First, we shortly describe how to apply likelihood-based boosting on the \texttt{bodyfat} data \cite{garcia}, included in \textbf{mboost}. The fitting functions are now called \R{GLMBoost()} and \R{GAMBoost()}, the number of stopping iteration can be specified via the \R{stepno} argument (as in \textbf{mboost}, the default is 100). The distribution for the likelihood is defined by the \R{family} argument. The predictors must be provided by a $n \times p$ matrix.

\begin{small}
\begin{verbatim}
R> library(GAMBoost) ## load package
R> library(mboost)   ## for the data
R> data(bodyfat)     ## load data
R>
R> glm1 <- GLMBoost(y = bodyfat$DEXfat,
+                   x = as.matrix(bodyfat[,c("hipcirc","kneebreadth","anthro3a")]),
+                   family = gaussian(), stepno = 100)
R> summary(glm1)
 family: gaussian (with canonical link)
 model components: 0 smooth, 3 linear (with penalty 71)
 model fit:
     at final boosting step (step 100):
         residual deviance 838.7505, df 3.99, AIC (corrected) 3.6227, BIC 860.0212
     at minimum AIC (corrected) (step 8):
         residual deviance 842.4226, df 2.72, AIC (corrected) 3.5859, BIC 858.2864
     at minimum BIC (step 10):
         residual deviance 839.1285, df 2.86, AIC (corrected) 3.5863, BIC 855.5622
 fitted covariates:
     at final boosting step (step 100):
         intercept: 30.7828
         3 non-zero estimates for linear terms:
         hipcirc (5.6476), kneebreadth (1.7377), anthro3a (4.2148)
     at minimum AIC (corrected) (step 8):
         intercept: 30.7828
         3 non-zero estimates for linear terms:
         hipcirc (5.6187), kneebreadth (1.7058), anthro3a (4.1334)
     at minimum BIC (step 10):
         intercept: 30.7828
         3 non-zero estimates for linear terms:
         hipcirc (5.6526), kneebreadth (1.7058), anthro3a (4.1933)
\end{verbatim}
\end{small}

The function \R{summary()} not only provides some information on the model fit and the coefficients but already includes the optimal stopping iteration regarding the AIC (corrected version by Hurvich et al. \cite{Hurvich98}) and the BIC.

The function \R{cv.GLMBoost} can be used to select the number of boosting steps by cross-validation:
\begin{small}
\begin{verbatim}
R> set.seed(123)
R> cv.glm1 <- cv.GLMBoost(y = bodyfat$DEXfat,
+              x = as.matrix(bodyfat[,c("hipcirc","kneebreadth","anthro3a")]),
+              family = gaussian(), maxstepno = 100)
R> cv.glm1$selected
 [1] 99
R> glm1 <- GLMBoost(y = bodyfat$DEXfat,
+            x = as.matrix(bodyfat[,c("hipcirc","kneebreadth","anthro3a")]),
+            family = gaussian(), stepno = cv.glm1$selected)
\end{verbatim}
\end{small}

For fitting generalized additive models, the equivalent \R{GAMBoost} functions can be used:
\begin{small}
\begin{verbatim}
R> set.seed(123)
R> cv.gam1 <- cv.GAMBoost(y = bodyfat$DEXfat,
+              x = as.matrix(bodyfat[,c("hipcirc","kneebreadth","anthro3a")]),
+              family = gaussian(), maxstepno = 100, just.criterion=TRUE)
R> cv.gam1$selected
 [1] 25
R> gam1 <- GAMBoost(y = bodyfat$DEXfat,
+              x = as.matrix(bodyfat[,c("hipcirc","kneebreadth","anthro3a")]),
+              family = gaussian(), stepno = cv.gam1$selected)
\end{verbatim}
\end{small}

The \R{plot} method provides the fitted functions together with approximate 95\% confidence intervals:
\begin{small}
\begin{verbatim}
R> par(mfrow=c(1,3))
R> plot(gam1)
\end{verbatim}
\end{small}
\includegraphics[width = \textwidth]{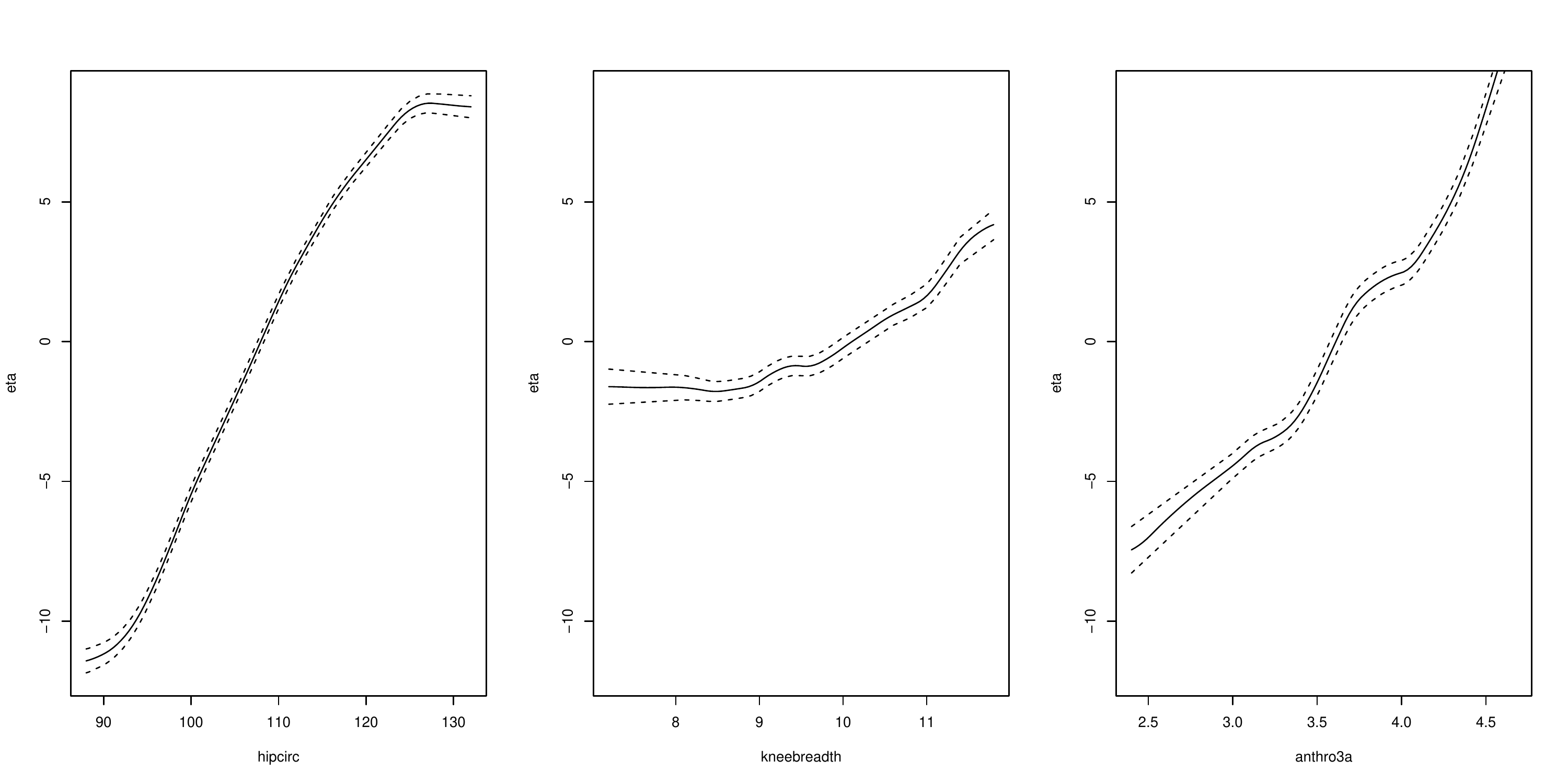}

\subsubsection*{Likelihood-based boosting for the Cox model: \textbf{CoxBoost}}

The \textbf{CoxBoost} package implements the likelihood-based boosting approach for the Cox proportional hazard model. We will apply \textbf{CoxBoost} for the prediction of survival after chemotherapy for diffuse large-B-cell lymphoma (DLBCL), based on a data set provided by Rosenwald et al. \cite{rosenwald2002}. The main outcome is a potentially censored survival time from 240 patients with a median follow up of 2.8 years. In total,  57\% of the patients died during that time. As possible molecular predictors, there are 7399 microarray features available. As clinical predictors, the International Prognostic Index (IPI) is available for 222 patients (compare to \cite{BinSch2008}). The data set can be downloaded from the Web:

\begin{small}
\begin{verbatim}

R> ## The DLBCL data of Rosenwald et al. (2002)
R> pat    <- read.delim("http://llmpp.nih.gov/DLBCL/DLBCL_patient_data_NEW.txt")
R> intens <- read.delim("http://llmpp.nih.gov/DLBCL/NEJM_Web_Fig1data")
\end{verbatim}
\end{small}

\subsubsection*{Pre-processing of the Rosenwald data set}

Before the data set can be applied, some pre-processing is necessary (including imputing missing values, e.g.,  by mean of nearest neighbours \cite{li2004partial}). These procedures are not boosting related and will not be explained in detail.

\begin{small}
\begin{verbatim}
R> ## Preprocessing of DLBCL data:
R> geneid <- intens$UNIQID
R> patid <- names(intens)[3:ncol(intens)]
R> exprmat <- t(intens[,3:ncol(intens)])
R> colnames(exprmat) <- paste("ID", geneid, sep="")
R> exprmat <- exprmat[substr(patid, nchar(patid)-8, nchar(patid)) == "untreated",]
R> patid <- rownames(exprmat)
R> exprmat <- exprmat[match(pat$"DLBCL.sample..LYM.number.",
+                     as.numeric(substr(patid, regexpr("LYM", patid)+3,
+                     regexpr("LYM", patid)+5))),]
R> train <- pat$"Analysis.Set" == "Training"
R>
R> ## -------------------------  LONG RUNTIME    ------------------------
R> ## impute missing values by mean of the eight nearest neighbours
R> ## (as described in Li & Gui, 2004)
R> ## own nearest neighbour code modelled after the Hastie&Tibshirani
R> ## procedure in the R package pamr
R> #
R> # new.exprmat <- exprmat
R> # for (i in 1:ncol(exprmat)) {
R> # cat(i," ")
R> # actual.na <- is.na(exprmat[,i])
R> # if (any(actual.na)) {
R> # neighbour.index <- sort(apply((exprmat - exprmat[,i])^2, 2,
R> #                         mean, na.rm = TRUE), index.return = TRUE)$ix
R> # new.exprmat[actual.na,i] <- apply(exprmat[actual.na, , drop = FALSE], 1,
R> #    function(arg) mean(arg[neighbour.index][!is.na(arg[neighbour.index])][1:8]))
R> # }}
R> # cat("\n")
R> # exprmat <- new.exprmat
R> ## --------------------------------------------------------------------
R>
R> ## simplistic median imputation (much faster)
R> exprmat <- ifelse(is.na(exprmat), matrix(apply(exprmat, 2, median, na.rm = TRUE),
+                    nrow(exprmat), ncol(exprmat), byrow = TRUE), exprmat)
R>
R> obs.time <- pat$"Follow.up..years."
R> obs.status <- ifelse(pat$"Status.at.follow.up" == "Dead", 1, 0)
R>
R> ## Use only patients, where IPI is available:
R> IPI <- pat$IPI.Group
R> IPI.available <- ifelse(is.na(IPI),FALSE,ifelse(IPI == "missing", FALSE , TRUE))
R> IPI <- IPI[IPI.available]
R> obs.time <- obs.time[IPI.available]
R> obs.status <- obs.status[IPI.available]
R> exprmat <- exprmat[IPI.available,]
\end{verbatim}
\end{small}

\subsubsection*{Model fitting}

First, we prepare the predictor matrix \R{xmat}, which contains both the molecular data and clinical predictors, and a vector \R{unpen.index} that indicates the position of the covariates that should receive unpenalized estimates (in this case the clinical predictors).

\begin{small}
\begin{verbatim}
R> IPI.medhigh <- ifelse(IPI != "Low",1,0)
R> IPI.high <- ifelse(IPI == "High",1,0)
R> xmat <- cbind(IPI.medhigh,IPI.high,exprmat)
R> colnames(xmat) <- c("IPImedhigh","IPIhigh",colnames(exprmat))
R> unpen.index <- c(1,2)
\end{verbatim}
\end{small}

Afterwards, we specify the penalty such as to obtain a step size of 0.1, similar to gradient boosting, as well as the number of boosting iterations:

\begin{small}
\begin{verbatim}
R> nu <- 0.1
R> penalty <- sum(obs.status)*(1/nu-1)
R> stepno <- 50
\end{verbatim}
\end{small}

Finally we fit a Cox model by the function \R{CoxBoost()} with and without adjusting for the IPI score.

\begin{small}
\begin{verbatim}
R> ## all predictors, including IPI: xmat
R> cb1.IPI <- CoxBoost(time = obs.time, status = obs.status, xmat,
+                      unpen.index = unpen.index, stepno = stepno,
+                      penalty = penalty, standardize = TRUE, trace = FALSE)
R>
R> ## only molecular data: exprmat
R> cb1 <- CoxBoost(time = obs.time, status = obs.status, exprmat,
+                  stepno = stepno, standardize = TRUE, trace = FALSE)
\end{verbatim}
\end{small}

Inspecting the selected genes
\begin{small}
\begin{verbatim}
R> summary(cb1.IPI)
50 boosting steps resulting in 26 non-zero coefficients (with 2 being mandatory)
partial log-likelihood: -558.12

Parameter estimates for mandatory covariates at boosting step 50:
           Estimate
IPImedhigh   0.9849
IPIhigh      0.8273

Optional covariates with non-zero coefficients at boosting step 50:
parameter estimate > 0:
 IPImedhigh, IPIhigh, ID31242, ID31981, ID34546, ID26474, ID34344, ID30931,
 ID24530, ID32302, ID32238, ID17385, ID33312
parameter estimate < 0:
 ID29871, ID27774, ID17154, ID29847, ID28325, ID24394, ID33157, ID19279,
 ID24376, ID16359, ID17726, ID20199, ID32679

R> summary(cb1)
50 boosting steps resulting in 28 non-zero coefficients
partial log-likelihood: -577.2544

Optional covariates with non-zero coefficients at boosting step 50:
parameter estimate > 0:
 ID24980, ID32424, ID28925, ID31242, ID34805, ID31981, ID31669, ID29176,
 ID17733, ID24400, ID34344, ID30634, ID29657, ID31254, ID33358, ID32238,
 ID34376
parameter estimate < 0:
 ID27774, ID24394, ID19279, ID24376, ID28641, ID27267, ID25977, ID29797,
 ID20199, ID28377, ID32679
\end{verbatim}
\end{small}

Now we can analyse the overlap between the two Cox models:

\begin{small}
\begin{verbatim}
R> intersect(names(coef(cb1.IPI)[coef(cb1.IPI) != 0]),
+            names(coef(cb1)[coef(cb1) != 0]))
 [1] "ID27774" "ID31242" "ID31981" "ID24394" "ID19279" "ID24376" "ID34344"
 [8] "ID32238" "ID20199" "ID32679"
 \end{verbatim}
\end{small}

The coefficient paths, i.e. parameter estimates plotted against the boosting steps, show a marked difference due to adjusting for the clinical covariates:
\begin{small}
\begin{verbatim}
R> par(mfrow=c(1,2))
R> plot(cb1)
R> plot(cb1.IPI)
\end{verbatim}
\end{small}
\begin{center}
\includegraphics[scale = 0.75]{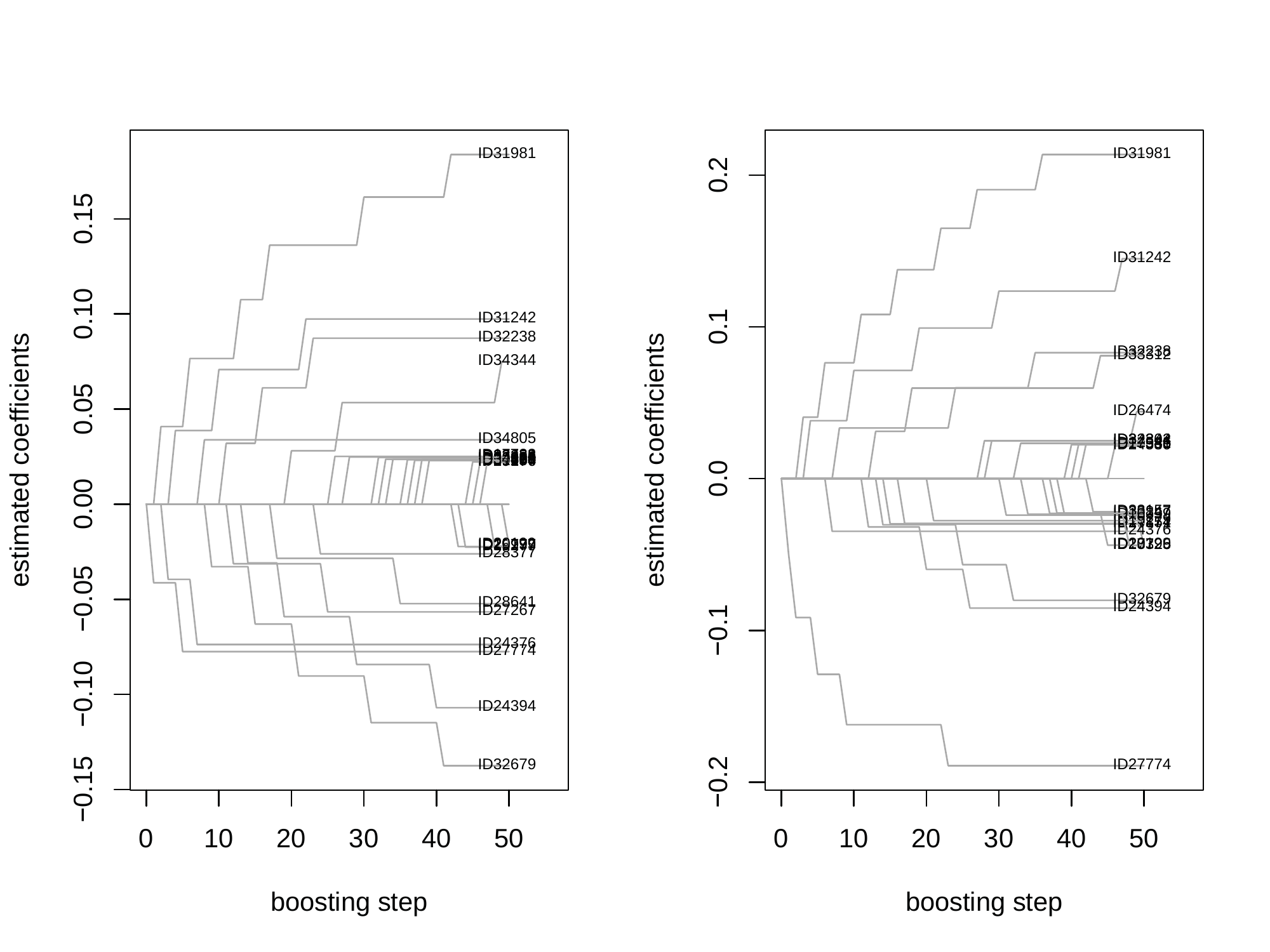}
\end{center}

\subsubsection*{Early stopping}

Similar to the packages \textbf{mboost} and \textbf{GAMBoost}, the \textbf{CoxBoost} package provides automated functionss to select the optimal stopping iteration. In our case, we will apply the function \R{cv.CoxBoost()} to determine the optimal value based on 10-fold cross-validation.

\begin{small}
\begin{verbatim}
R> set.seed(123)
R> cv1 <- cv.CoxBoost(time = obs.time, status = obs.status, x = xmat,
+                     unpen.index = unpen.index, maxstepno = 200, K = 10,
+                     standardize = TRUE, trace = TRUE, penalty = penalty,
+                     multicore = FALSE)
\end{verbatim}
\end{small}

If the package \R{parallel} is available, one can specify \R{multicore = TRUE} leading to automated parallel computing (if the machine contains more than one core).

The optimal value is then the one optimizing the average likelihood on the 10 test folds:

\begin{small}
\begin{verbatim}
R> cv1$optimal.step
[1] 35
\end{verbatim}
\end{small}

Now we can compute our final Cox model. The \R{trace} option provides a live view of the gene selection in the boosting steps:

\begin{small}
\begin{verbatim}
R> cb1.IPI.cv <- CoxBoost(time = obs.time, status = obs.status, xmat,
+                         unpen.index = unpen.index, stepno = cv1$optimal.step,
+                         penalty = penalty, standardize = TRUE, trace = TRUE)
 ID27774 ID27774 ID31981 ID31242 ID27774 ID31981 ID24376 ID32238
 ID27774 ID31242 ID31981 ID24394 ID33312 ID32679 ID29871 ID31981
 ID17154 ID33312 ID31242 ID24394 ID19279 ID31981 ID27774 ID32238
 ID32679 ID24394 ID31981 ID32302 ID34344 ID31242 ID16359 ID32679
 ID26474 ID28325 ID32238
\end{verbatim}
\end{small}

\end{document}